\documentclass[nojss]{jss}
\usepackage{verbatim}
\usepackage{amssymb,amsmath}
\usepackage{natbib}
\usepackage{color,soul}
\graphicspath{{./Figures/}}
\DeclareMathOperator*{\argmax}{arg\,max}
\author{Agnieszka Kr\'{o}l\\ISPED, INSERM U897\\Universit\'e de Bordeaux  \And Audrey Mauguen\\ISPED, INSERM U897\\Universit\'e de Bordeaux  \And
Yassin Mazroui\\Universit\'{e} Pierre et Marie Curie\\\& INSERM, UMR S 1136, Paris \AND
Alexandre Laurent\\ISPED, INSERM U897\\Universit\'e de Bordeaux \And
Stefan Michiels \\ INSERM U1018, CESP\\ \& Gustave Roussy, Villejuif \And
Virginie Rondeau\\ISPED, INSERM U897\\Universit\'e de Bordeaux}
\title{Tutorial in Joint Modeling and Prediction: a Statistical Software for Correlated Longitudinal Outcomes, Recurrent Events and a Terminal Event}
\Plainauthor{Agnieszka Kr\'{o}l, Audrey Mauguen, Yassin Mazroui, Alexandre Laurent, Stefan Michiels, Virginie Rondeau} 
\Plaintitle{Tutorial in Joint Modeling and Prediction: a statistical software for correlated longitudinal outcomes, recurrent events and a terminal event}
\Shorttitle{Joint Modeling and Prediction} 

\Abstract{Extensions in the field of joint modeling of correlated data and dynamic predictions improve the development of prognosis research. The \proglang{R} package \pkg{frailtypack} provides estimations of various joint models for longitudinal data and survival events. In particular, it fits models for recurrent events and a terminal event (\code{frailtyPenal}), models for two survival outcomes for clustered data (\code{frailtyPenal}), models for two types of recurrent events and a terminal event (\code{multivPenal}), models for a longitudinal biomarker and a terminal event (\code{longiPenal}) and models for a longitudinal biomarker, recurrent events and a terminal event (\code{trivPenal}). The estimators are obtained using a standard and penalized maximum likelihood approach, each model function allows to evaluate goodness-of-fit analyses and plots of baseline hazard functions. Finally, the package provides individual dynamic predictions of the terminal event and evaluation of predictive accuracy. This paper presents theoretical models with estimation techniques, applies the methods for predictions and illustrates \pkg{frailtypack} functions details with examples.}

\Keywords{dynamic prediction, frailty, joint model, longitudinal data, predictive accuracy, \proglang{R}, survival analysis}
\Plainkeywords{}
\Address{

	Agnieszka Kr\'{o}l, Audrey Mauguen, Virginie Rondeau \\
	INSERM U1219 (Biostatistics Team), Universit\'e de Bordeaux \\
	146, rue L\'eo Saignat, 33076 Bordeaux Cedex, France\\
	Telephone: +33/5/57/57/45/31 \\
	E-mail: Agnieszka.Krol@isped.u-bordeaux2.fr, \\
	Audrey.Mauguen@isped.u-bordeaux2.fr\\
	 Virginie.Rondeau@isped.u-bordeaux2.fr \\
	URL: \url{http://www.bordeaux-population-health.center}\\
	\\
	Yassin Mazroui\\
	Sorbonne Universit\'{e}s, Universit\'{e} Pierre et Marie Curie,\\
	Laboratoire de Statistique Th\'{e}orique et Appliqu\'{e}e\\ 
	F-75005, Paris, France\\
	\textit{and}\\
    INSERM, UMR S 1136, \\
    Institut Pierre Louis d'\'{E}pid\'{e}miologie et de Sant\'{e} Publique\\
    F-75013, Paris, France \\
    E-mail: Yassin.Mazroui@upmc.fr\\
	\\
	Stefan Michiels\\
	Service de Biostatistique et d'Epidemiologie\\
	Gustave Roussy\\
	114 Rue Edouard Vaillant\\
	94800 Villejuif, France\\
	\textit{and}\\
	INSERM U1018, CESP\\
	Universit\'e Paris-Sud, Universit\'e Paris-Saclay\\
	Villejuif, France\\
	E-mail: Stefan.Michiels@gustaveroussy.fr
	}
\begin{document}
\sloppy
			\section{Introduction}
\subsubsection{Joint models}
Recent technologies allow registration of greater and greater amount of data. In the medical research different kinds of patient information are gathered over time together with clinical outcome data such as overall survival (OS). Joint models enable the analysis of correlated data of different types such as individual repeated data, clustered data together with OS. The repeated data may be recurrent events (e.g., relapses of a tumor, hospitalizations, appearance of new lesions) or a longitudinal outcome called biomarker (e.g., tumor size, prostate-specific antigen or CD4 cell counts). The correlated data that are not analyzed jointly with OS, are subjugated to a bias coming from ignoring the terminal event which is the competing event for the occurrence of repeated outcomes (not only it precludes the outcomes from being observed but also prevents them from occurring). On the other hand, the standard survival analysis for OS may lead to biased estimations, if the repeated data is considered as time-varying covariates or if it is completely ignored in the analysis.

For these correlated data one can consider joint models, e.g., a joint model for a longitudinal biomarker and a terminal event, which received the most of the attention in the literature. This joint model estimates simultaneously the longitudinal and survival processes using the relationship via a latent structure of random-effects \citep{Wulfsohn1997}. Extensions of these include, among others, models for multiple longitudinal outcomes \citep{Hatfield2012}, multiple failure times \citep{Elashoff2008} and both \citep{Chi2006}. A review of the joint modeling of longitudinal and survival data was already given elsewhere \citep{McCrink2013,LawrenceGould2014,Asar2015}. 

Another option for the analysis of correlated data are joint models for recurrent events and a terminal event \citep{Liu2004,Rondeau2007joint}. These models are usually called joint frailty models as the processes are linked via a random effect that represents the frailty of a subject (patient) to experience an event. They account for the dependence between two survival processes quantified by the frailty term, contrary to the alternative marginal approach \citep{Li1997}, which do not model the dependence. Some extensions to joint frailty models include incorporation of a nonparametric covariate function \citep{Yu2011}, inclusion of two frailty terms for the identification of the origin of the dependence between the processes \citep{Mazroui2012}, consideration of the disease-specific mortality process \citep{Belot2014} and accommodation of time-varying coefficients \citep{Yu2014,Mazroui2015}. Finally, \citet{Mazroui2013} proposed a model with two types of recurrent events following the approach of \citet{Zhao2012}. A review of joint frailty models in the Bayesian context was given by \citet{Sinha2008}.

Joint models for recurrent events and longitudinal data have received the least attention among the joint models so far. However, a marginal model based on the generalized linear mixed model was proposed by \citet{Efendi2013}. This model allows several longitudinal and time-to-event outcomes and includes two sets of random effects, one for the correlation within a process and between the processes, and the second to accommodate for overdispersion.

Finally, in some applications all the types of data, i.e., a longitudinal biomarker, recurrent events and a terminal event can be studied so that all of them are correlated to each other (in the following we call such models trivariate models). \citet{Liu2008} proposed a trivariate model for medical cost data. The longitudinal outcomes (amount of medical costs) were reported at informative times of recurrent events (hospitalizations) and were related to a terminal event (death). The dependence via random effects was introduced so that the process of longitudinal measurements and the process of recurrent events were related to the process of the terminal event. A relationship between longitudinal outcomes and  recurrent events was not considered. This relationship was incorporated to a trivariate  model proposed by \citet{Liu2009a} for an application of an HIV study. In this parametric approach the focus was on the analysis of the associations present in the model and the effect of the repeated measures and recurrent events on the terminal event.  \citet{Kim2012} analyzed the trivariate data using the transformation functions for the cumulative intensities for recurrent and terminal events. Finally, \citet{Krol2015} proposed a trivariate model in which all the processes were related to each other via a latent structure of random effects and applied the model to the context of a cancer evolution and OS.

\subsubsection{Prediction in joint models}
In the framework of joint models that consider a terminal event, one can be interested in predictions of the event derived from the model. As joint models include information from repeated outcomes, these predictions are dynamic, they change with the update of the observations. Dynamic predictive tools were proposed for joint models for longitudinal data and a terminal event \citep{Proust-Lima2009,Rizopoulos2011a}, joint models for  recurrent events and a terminal event \citep{Mauguen2013} and trivariate models \citep{Krol2015}. 

For the evaluation of the predictive accuracy of a joint model few methods were proposed due to the complexity of the models in which the survival data are usually right-censored. The standard methods for the assessment of the predictive abilities derived from the survival analysis and adapted to the joint models context are the Brier Score \citep{Proust-Lima2009,Mauguen2013,Krol2015} and receiver operating characteristic (ROC) methodology \citep{Rizopoulos2011a,Blanche2015}. Finally, the expected prognostic cross-entropy (EPOCE), a measure based on the information theory, provides a useful tool for the evaluation of predictive abilities of a model \citep{Commenges2012,Proust-Lima2014}.

\subsubsection{Software for joint models}
Together with the theoretical development of the joint models, the increase of appropriate software is observed, mostly for standard frameworks.  Among the \proglang{R} \citep{R2016} packages, the joint analysis of a single longitudinal outcome and a terminal event can be performed using \pkg{JM} \citep{JM} based on the likelihood maximization approach using expectation-maximization (EM) algorithm, \pkg{JMbayes} \citep{JMbayes} built in the Bayesian context, and \pkg{joineR} \citep{joineR}, a frequentist approach that allows a flexible formulation of the dependence structure. For the approach based on the joint latent class models, i.e., joint models that consider homogeneous latent subgroups of individuals sharing the same biomarker trajectory and risk of the terminal event, an extensive package \pkg{lccm} \citep{lcmm} provides the estimations based on the frequentist approach. Apart from the \proglang{R} packages, a \pkg{stjm} package \citep{stjm} in \proglang{STATA} uses maximum likelihood and provides flexible methods for modeling the longitudinal outcome based on polynomials or splines. Another possibility for the analysis of the joint models for a longitudinal outcome and a terminal event is a \proglang{SAS} macro \pkg{JMFit} \citep{JMFit}. Joint models using nonlinear mixed-effects models can be estimated using \proglang{MONOLIX} software \citep{Mbogning2015}. Among the packages \pkg{JM}, \pkg{JMbayes} and \pkg{lcmm} provide predictive tools and  predictive accuracy measures: EPOCE estimator in \pkg{lcmm}, ROC and AUC (area under the curve) measures and prediction error in \pkg{JM} and \pkg{JMbayes}. 

For joint models for correlated events (recurrent event or clustered event data) and a terminal event the available statistical software is limited. Among the \proglang{R} packages, \pkg{joint.Cox} \citep{joint.Cox}  provides the estimations  using penalized likelihood under the copula-based joint Cox models for time to clustered events (progression) and a terminal event. Trivariate joint models proposed by \citet{Liu2009a} were performed in \proglang{SAS} using the procedure NLMIXED.

The \proglang{R} package \pkg{frailtypack} \citep{frailtypack} fits several types of joint models. Originally developed for the shared frailty models for correlated outcomes it has been extended into the direction of joint models. These extensions include the joint model for one or two types of recurrent events and a terminal event, the joint model for a longitudinal biomarker and a terminal event and the trivariate model for a longitudinal biomarker, recurrent events and a terminal event. Moreover, \pkg{frailtypack} includes prediction tools, such as concordance measures adapted to shared frailty models and predicted probability of events for the joint models.  

Characteristics of a previous version of the package, such as estimation of several shared and standard joint frailty models using penalized likelihood, were already described elsewhere \citep{Rondeau2012frailtypack}. This work focuses on an overview and developments of joint models included in the package (model for recurrent/clustered events and a terminal event, model for multivariate recurrent events and a terminal event, model for longitudinal data and a terminal event and model for longitudinal data, recurrent events and a terminal event) and the prediction tools accompanied by predictive accuracy measures. Finally, several options available for the models (e.g., two correlated frailties in the model for recurrent events and a terminal event, left-censoring for the longitudinal data) will be presented with the appropriate examples. A practical guide to different types of models included in the package along with available options is presented in a schematic table in Appendix A. 
 
This article firstly presents joint models with details on estimation methods (Section ~\ref{joint_models}) and predictions of an event in the framework of these joint models (Section~\ref{sec:prediction}). In Section~\ref{package}, the \pkg{frailtypack} package features are detailed. Section~\ref{examples} contains some examples on real datasets and finally, Section~\ref{conclusion} concludes the paper.

\section{Models for correlated outcomes}
\label{joint_models}
\subsection{Bivariate joint frailty models for two types of survival outcomes}
\subsubsection{a) Joint model for recurrent events and terminal event}
We denote by $i=1,\ldots,N$ the subject and by $j=1,\ldots,n_i$ the rank of the recurrent event. For each subject $i$ we observe the time of the terminal event $T_i=\textrm{min}(T_i^*,C_i)$ with $T_i^*$ the true terminal event time and $C_i$ the censoring time. We denote the observed recurrent times $T_{ij}=\textrm{min}(T_{ij}^*,C_i,T_i^*)$ with $T_{ij}^*$ the real time of  recurrent event. Thus, for each rank $j$ we can summarize the information with a triple $\{T_{ij},\delta_{ij},\delta_i\}$, where $\delta_{ij}=I_{\{T_{ij}=T_{ij}^*\}}$ and $\delta_i=I_{\{T_i=T_i^*\}}$. Let $r_0(.)$ and $\lambda_0(.)$ be baseline hazard functions related to the recurrent and terminal events, respectively. Let $\textbf{X}_{Rij}$ and $\textbf{X}_{Ti}$ be two vectors of covariates associated with the risk of the related events. Let  $\boldsymbol\beta_R$  and $\boldsymbol\beta_T$ be constant effects of the covariates whereas $\boldsymbol\beta_R(t)$  and $\boldsymbol\beta_T(t)$ denote a time-dependent effect. Finally, a frailty $u_i$ is a random effect that links the two risk functions and follows a given distribution $\mathcal{D}$ and $\alpha$ is a parameter allowing more flexibility. The hazard functions are defined by \citep{Liu2004,Rondeau2007joint}:
\begin{equation}
\label{jointrecsurv}
\left\lbrace
\begin{array}{lc}
  r_{ij}(t|u_i)=u_ir_0(t)\exp({{\boldsymbol X_{Rij}^\top \boldsymbol \beta_R}})= u_i r_{ij}(t)& \textrm{(recurrent event)} \\
\lambda_i(t|u_i)=u_i^{\alpha}\lambda_0(t)\exp({{\boldsymbol X_{Ti}^\top \boldsymbol \beta_T}})= u_i^{\alpha}\lambda_i(t) & \textrm{(terminal event)}
\end{array},
\right.
\end{equation}
where the frailty terms $u_i$ are $iid$ (independent and identically distributed) and follow either the Gamma distribution with unit mean and variance $\theta$ ($\theta>0$), i.e., $ u_i \sim \mathcal{G}(\frac{1}{\theta},\frac{1}{\theta}) $, or the log-normal distribution, i.e., $u_i= \exp(v_i) \sim \textrm{ln}\mathcal{N}(0,\sigma_v^2) $. The parameter $\alpha$ determines direction of the association (if found significant) between the processes. 

For a given subject we can summarize all the information with $\boldsymbol \Theta_i=\{\boldsymbol T_i^{(1)},T_i,\boldsymbol \delta_i^{(1)},\delta_i\}$, where $\boldsymbol T_i^{(1)}=\{T_{ij},j=1,\ldots,n_i\}$ and $\boldsymbol \delta_i^{(1)}=\{\delta_{ij},j=1,\ldots,n_i\}$. Finally, we are interested to estimate $\boldsymbol \xi=\{r_0(\cdot),\lambda_0(\cdot),\boldsymbol\beta_R,\boldsymbol\beta_T,\theta,\alpha\}$.

In some cases, e.g.,  long follow-up, effects of certain prognostic factors can be varying in time. For this reason a joint model with time-dependent coefficients was proposed by \citet{Mazroui2015}. The coefficients $\boldsymbol \beta_R(t)$ and $\boldsymbol \beta_T(t)$ are modeled with regression B-splines of order $q$ and $m$ interior knots. The general form of an estimated time-dependent coefficient $\widehat{\beta(t)}$ is
$$ \widehat{\beta(t)}=\sum_{j=-q+1}^{m}\widehat{\zeta_j}B_{j,q}(t),$$
where $B_{j,q}(t)$ is the basis of B-splines calculated using a recurring expression \citep{DeBoor2001}. Therefore, for every time-varying coefficient, $m+q$ parameters $\zeta_j$, $j=-q+1,\ldots,m$ need to be estimated. It has been shown that quadratic B-splines, i.e., $q=3$, with a small number of interior knots ($m\leq 5$) ensure stable estimation \citep{Mazroui2015}. The pointwise confidence intervals for $\beta(t)$ can be obtained using the Hessian matrix.

The application of time-varying coefficients is helpful if there is a need to verify the proportional hazard (PH) assumption. Using a likelihood ratio (LR) test, the time-dependency of a covariate can be examined. A model with the time-dependent effect and a model with the constant effect are compared and the  test hypotheses depend on whether the covariate is related to one of the events or both. If the time-dependency is tested for both events the null hypothesis is $H_0\ :\ \beta_R(t)=\beta_R,\ \beta_T(t)=\beta_T$ and the alternative hypothesis is $H_1\ :\ \beta_R(t)\ne\beta_R,\ \beta_T(t)\ne\beta_T$. The LR statistic has a $\chi^2$ distribution of degree $k(m+q-1)$, where $k$ is the number of events to which the covariate is related.

The LR test can also be used to verify whether a covariate with the time-dependent effect is significant for  the events. In this case, a model with the time-varying effect covariate is compared to a model without this covariate. Again, the null hypotheses depend on the events considered. If the covariate is associated to both events, the null hypothesis $H_0\ :\ \beta_R(t)=0,\ \beta_T(t)=0$ is against $H_1\ :\ \beta_R(t)\ne 0,\ \beta_T(t)\ne 0$. The LR statistic has a $\chi^2$ distribution of degree $k(m+q)$.

\subsubsection{b) Joint model for two clustered survival events}
An increasing number of studies favor the presence of clustered data, especially multi-center or multi-trial studies. In particular, meta-analyses are used to calculate surrogate endpoints in oncology. However, the clustering creates some heterogeneity between subjects, which needs to be accounted for.  Using the above notations, the clustered joint model is similar to the model (\ref{jointrecsurv}) but here, the index $j$ ($j=1,\ldots,n_i$) represents a subject from cluster $i$ ($i=1,\ldots,N$) and the cluster-specific frailty term $u_i$ is shared by the subjects of a given group. Thus, the model can be written as \citep{Rondeau2011joint}:
\begin{equation}
\label{jointcluster}
\left\lbrace
\begin{array}{lc}
  r_{ij}(t|u_i)=u_ir_0(t)\exp({{\boldsymbol X_{Rij}^\top \boldsymbol \beta_R}})= u_i r_{ij}(t)& \textrm{(Time to event 1)} \nonumber\\
\lambda_{ij}(t|u_i)=u_i^{\alpha}\lambda_0(t)\exp({{\boldsymbol X_{Tij}^\top \boldsymbol \beta_T}})= u_i^{\alpha}\lambda_{ij}(t) & \textrm{(Time to event 2)}
\end{array},
\right.
\end{equation}
and we assume $iid$ the Gamma distribution for the frailty terms $u_i$. The events can be chosen arbitrarily but it is assumed that the event 2 impedes the process of the event 1. Usually the event 2 is death of a patient and the other is an event of interest (e.g., surrogate endpoint for OS) such as time to tumor progression or progression-free survival.

The interest of using the joint model for the clustered data is that it considers the dependency between the survival processes and respects that event 2 is a competitive event for event 1. The frailty term $u_i$ is common for a given group and represents the clustered association between the processes (at the cluster level) as well as the intra-cluster correlation.

The package \pkg{frailtypack} includes clustered joint models for two survival outcomes in presence of semi-competing risks (time-to-event, recurrent events are not allowed). To ensure identifiability of the models, it is assumed that $\alpha=1$. For these models, only Gamma distribution of the frailty term is implemented in the package.

\subsubsection{c) Joint model for recurrent events and a terminal event with two frailty terms}
In Model (\ref{jointrecsurv}) the frailty term $u_i$ reflects the inter- and intra-subject correlation for the recurrent event as well as the association between the recurrent and the terminal events. In order to distinguish the origin of dependence, two independent frailty terms $\boldsymbol u_i=(u_i$,$v_i)$ can be considered \citep{Mazroui2012}:
\begin{equation}
\label{jointgeneral}
\left\lbrace
\begin{array}{lc}
  r_{ij}(t|u_i)=u_i v_i r_0(t)\exp({{\boldsymbol X_{Rij}^\top \boldsymbol \beta_R}})= u_i v_i r_{ij}(t)& \textrm{(recurrent event)} \\
\lambda_i(t|u_i)=u_i\lambda_0(t)\exp({{\boldsymbol X_{Ti}^\top \boldsymbol \beta_T}})= u_i\lambda_i(t) & \textrm{(terminal event)}
\end{array},
\right.
\end{equation}
where $v_i\sim\Gamma(\frac{1}{\eta},\frac{1}{\eta})$ ($\eta>0$) specific to the recurrent event rate and $u_i\sim\Gamma(\frac{1}{\theta},\frac{1}{\theta})$ ($\theta>0$) specific to the association between the processes. The variance of the frailty terms represent the heterogeneity in the data, associated with unobserved covariates. Moreover, high value of the variance $\eta$ indicates strong dependence between the recurrent events and high value of the variance $\theta$ indicates that the recurrent and the terminal events are strongly dependent. The individual information is equivalent to $\boldsymbol \Theta_i$ from Model (\ref{jointrecsurv}) and the parameters to estimate are  $\boldsymbol \xi=\{r_0(\cdot),\lambda_0(\cdot),\boldsymbol\beta_R,\boldsymbol\beta_T,\theta,\eta\}$.

In the package \pkg{frailtypack} for Model (2), called also general joint frailty model, only Gamma distribution is allowed for the random effects.

\subsection{Multivariate joint frailty model}
One of extensions to the joint model for recurrent events and a terminal event (\ref{jointrecsurv}) is consideration of different types of recurrence processes and a time-to-event data. Two types of recurrent events are taken into account in a multivariate frailty model proposed by \citet{Mazroui2013}. The aim of the model is to analyze dependencies between all types of events. The recurrent event times are defined by $T_{ij}^{(1)}$ ($j=1,\ldots,n_i^{(1)}$) and $T_{ij}^{(2)}$ ($j=1,\ldots,n_i^{(2)}$) and both processes are censored by the terminal event $T_i$. The joint model is expressed using the recurrent events and terminal event hazard functions:
\begin{equation}
\label{joint2recsurv}
\left\lbrace
\begin{array}{lc}
  r^{(1)}_{ij}(t|u_i,v_i)=r_0^{(1)}(t)\exp({u_i+{\boldsymbol X_{Rij}^{(1)}}^\top \boldsymbol \beta_{R}^{(1)}})= \exp{(u_i)} r_{ij}^{(1)}(t)& \textrm{(rec. event 1)} \\
    r^{(2)}_{ij}(t|u_i,v_i)=r_0^{(2)}(t)\exp({v_i+{\boldsymbol X_{Rij}^{(2)}}^\top \boldsymbol \beta_{R}^{(2)}})= \exp({v_i}) r_{ij}^{(2)}(t)& \textrm{(rec. event 2)} \\
\lambda_{i}(t|u_i,v_i)=\lambda_0(t)\exp({\alpha_1 u_i+\alpha_2 v_i+{\boldsymbol X_{Ti}^\top \boldsymbol \beta_T}})= \exp({\alpha_1 u_i+\alpha_2 v_i)}\lambda_{i}(t) & \textrm{(terminal event)}
\end{array},
\right.
\end{equation}
with vectors of regression coefficients $\boldsymbol \beta_R^{(1)},\boldsymbol \beta_R^{(2)},\boldsymbol \beta_T$ and covariates $\boldsymbol X_{Rij}^{(1)},\boldsymbol X_{Rij}^{(2)},\boldsymbol X_{Ti}$ for the first recurrent event, the second recurrent event and the terminal event, respectively. The frailty terms $\boldsymbol u_i=(u_i,v_i)^\top$ explain the intra-correlation of the processes and the dependencies between them. For these correlated random effects the multivariate normal distribution of dimension 2 is considered:
 
\begin{equation}
 \boldsymbol u_i=\begin{pmatrix} u_i\\v_i \\ \end{pmatrix} \sim \mathcal{N}\left(\textbf{0},
 \begin{pmatrix}   \sigma_u^2 & \rho\sigma_u \sigma_v \\
     \rho\sigma_u \sigma_v &  \sigma_v^2\\\end{pmatrix}\right),\nonumber
 \end{equation}
Variances $\sigma_u^2$ and $\sigma_v^2$ explain the within-subject correlation between occurrences of the recurrent event of type 1 and type 2, respectively. The dependency between the two recurrent events is explained by the correlation coefficient $\rho$ and the dependency between the recurrent event $1$ (event 2) and the terminal event is assessed by the term $\alpha_1$ ($\alpha_2$) in case of significant variance $\sigma_u^2$ ($\sigma_v^2$).
 
Here, for each subject $i$ we observe $\boldsymbol \Theta_i=\{\boldsymbol T_i^{(1)},\boldsymbol T_i^{(2)},T_i,\boldsymbol \delta_i^{(1)},\boldsymbol \delta_i^{(2)},\delta_i\}$, where $\boldsymbol T_i^{(l)}=\{T_{ij}^{(l)},j=1,\ldots,n_i^{(l)}\}$ and $\boldsymbol \delta_i^{(l)}=\{\delta^{(l)}_{ij},j=1,\ldots,n_i^{(l)}\}$, $l=1,2$ and the parameters to estimate are $\boldsymbol \xi = \{r_0^{(1)}(\cdot),r_0^{(2)}(\cdot),\lambda_0(\cdot),\boldsymbol\beta_R^{(1)},\boldsymbol\beta_R^{(2)},\boldsymbol\beta_T,\sigma_{u}^2,\sigma_v^2,\rho,\alpha_1,\alpha_2\}$.

\subsection{Bivariate joint model with longitudinal data}
Here, instead of recurrent events we consider a longitudinal biomarker. For subject $i$ we observe an  $l_i$-vector of longitudinal measurements $\boldsymbol y_i=\{y_i(t_{ik}),k=1,\ldots,l_{i}\}$. Again, the true terminal event time $T_i^*$ impedes the longitudinal process and the censoring time does not stop it but values of the biomarker are no longer observed. For each subject $i$ we observe $\boldsymbol \Theta_i=\{\boldsymbol y_i,T_i,\delta_i\}$. A random variable $ Y_i(t)$ representing the biomarker is expressed using a linear mixed-effects model and the terminal event time $T_i^*$ using a proportional hazards model. The sub-models are linked by the random effects $\boldsymbol u_i = \boldsymbol b_i^\top$:
\begin{equation}
\label{srem}
\left\{  \begin{array}{lc}
   Y_{i}(t)= m_i(t)+ \epsilon_i(t)={\boldsymbol X_{Li}(t)}^\top  \boldsymbol \beta_L + {\boldsymbol Z_i(t)}^\top  \boldsymbol b_i +  \epsilon_i(t)& \textrm{(biomarker)}\\
  \lambda_i(t|\boldsymbol b_i)=\lambda_{0}(t)\ \exp({{\boldsymbol X_{Ti}}^\top \boldsymbol \beta_T+h(\boldsymbol b_i,\boldsymbol \beta_L,\boldsymbol Z_{i}(t),\boldsymbol X_{Li}(t))^\top \boldsymbol \eta_T })& \textrm{(death)}\\
   \end{array},
\right. 
\end{equation}
where $\boldsymbol X_{Li}(t)$ and $\boldsymbol X_{Ti}$ are vectors of fixed effects covariates. Coefficients $\boldsymbol \beta_L$ and $\boldsymbol \beta_T$ are constant in time. Measurements errors $\epsilon_i(\cdot)$ are $iid$ normally distributed with mean 0 and variance $\sigma_{\epsilon}^2$. We consider a $q$-vector of the random effects parameters $\boldsymbol b_i=(b_0,\ldots,b_{q-1})^\top\sim\mathcal{N}(\textbf{0},\textbf{B}_1) $ associated to  covariates $\boldsymbol Z_{i}(t)$ and independent from the measurement error. In the \pkg{frailtypack} package, the maximum size of the matrix $\boldsymbol B_1$ is 2. The relationship between the two processes is explained via a function $h(\boldsymbol b_i,\boldsymbol \beta_L,\boldsymbol Z_{i}(t),\boldsymbol X_{Li}(t))$ and quantified by coefficients $\boldsymbol \eta_T$. This possibly multivariate function represents the prognostic structure of the biomarker $ m_i(t)$ (we assume that the measurement errors are not prognostic for the survival process) and in the \pkg{frailtypack} these are either the random effects $\boldsymbol b_i$ or the current biomarker level $ m_i(t)$. The structure of dependence is chosen \textit{a priori} and should be designated with caution as it influences the model in terms of fit and predictive accuracy.

We consider that the longitudinal outcome $y_i(t_{ik})$ can be a subject to a quantification limit, i.e., some observations, below a level of detection $s$ cannot be quantified (left-censoring). We introduce it in the vector $\boldsymbol y_i$ which includes $l_i^o$ observed values $\boldsymbol y_i^o$ and $l_i^c$ censored values $\boldsymbol y_i^c$ ($l_i=l_i^o+l_i^c$). The aspect of the left-censored data is handled in the individual contribution from the longitudinal outcome to the likelihood. The individual observed outcomes are $\boldsymbol \Theta_i=\{\boldsymbol y_i^o,\boldsymbol y_i^c,T_i,\delta_i\}$ (in case of no left-censoring $\boldsymbol y_i^c$ is empty and $\boldsymbol y_i^o$ is equivalent to $\boldsymbol y_i$) and the parameters to estimate are $\boldsymbol \xi=\{\lambda_0(\cdot),\boldsymbol\beta_L,\boldsymbol\beta_T,\boldsymbol B_1,\sigma_{\epsilon}^2,\boldsymbol \eta_T\}$.
 
\subsection{Trivariate joint model with longitudinal data}
The trivariate joint model combines the joint model for recurrent events and a terminal event (Model (\ref{jointrecsurv})) with the joint model for a longitudinal data and a terminal event (\ref{srem}). We consider the longitudinal outcome $Y$ observed in discrete time points $\boldsymbol y_i=\{y_i(t_{ik}),k=1,\ldots,l_{i}\}$, times of the recurrent event $T_{ij}$ ($j=1,\ldots,n_i$) and the terminal event time $T_i$ that is an informative censoring for the longitudinal data and recurrences.
The respective processes are linked to each other via a latent structure. We define multivariate functions $g(\cdot)$ and $h(\cdot)$ for the associations between the biomarker and the recurrent events and between the biomarker and the terminal event, respectively. For the link between the recurrent events and the
terminal event we use a frailty term $v_i$ from the normal distribution. This distribution was chosen here to facilitate the estimation procedure involving multiple numerical integration. Interpretation of $v_i$ should be performed with caution as the dependence between the recurrences and the terminal data is partially explained by the random-effects of the biomarker trajectory. We define the model by \citep{Krol2015}:
\begin{equation}
\label{trivariate}
\left\{  \begin{array}{lc}
 Y_i(t)= m_i(t)+\epsilon_i(t)={\boldsymbol X_{Li}(t)}^\top \boldsymbol \beta_L + \boldsymbol Z_{i}(t)^\top  \boldsymbol b_{i} +   \epsilon_i(t)& \textrm{(biomarker)}\\
   r_{ij}(t|v_i,\boldsymbol b_{i})=r_0(t)\ \exp({v_i+{\boldsymbol X_{Rij}}^\top \boldsymbol \beta_R+g(\boldsymbol b_{i},\boldsymbol \beta_L,\boldsymbol Z_{i}(t),\boldsymbol X_{Li}(t))^\top \boldsymbol \eta_{R}})& \textrm{(recurrent event)}\\
  \lambda_i(t|v_i,\boldsymbol b_{i})=\lambda_{0}(t)\ \exp({\alpha  v_i+{\boldsymbol X_{Ti}}^\top \boldsymbol \beta_T+h(\boldsymbol b_{i},\boldsymbol \beta_L,\boldsymbol Z_{i}(t),\boldsymbol X_{Li}(t))^\top \boldsymbol \eta_{T}})& \textrm{(terminal event)}\\
   \end{array}
\right. ,
\end{equation}
where the regression coefficients $\boldsymbol \beta_L,\boldsymbol \beta_R,\boldsymbol \beta_T$ are associated to possibly time-varying covariates $\boldsymbol X_{Li}(t), \boldsymbol X_{Rij},\boldsymbol X_{Ti}$ for the biomarker, recurrent events and the terminal event (baseline prognostic factors), respectively. The strength of the dependencies between the processes is quantified by $\alpha$ and vectors $\boldsymbol \eta_{R}$ and $\boldsymbol \eta_T$. The vector of the random effects $\boldsymbol u_i=(\boldsymbol b_{i}^\top,v_i)^\top $ of dimension $q=q_1+1$ follows the multivariate normal distribution:
\begin{equation}
 \boldsymbol u_i=\begin{pmatrix} \boldsymbol b_{i}\\v_i \\ \end{pmatrix} \sim \mathcal{N}\left(\textbf{0},
 \begin{pmatrix}   \textbf{B}_1&\textbf{0} \\
      \textbf{0}  &  \sigma_v^{2}\\\end{pmatrix}\right),\nonumber
 \end{equation}
where the dimension of the matrix $\textbf{B}_1$ in the \pkg{frailtypack} package is allowed to be maximum 2. It is convenient if one assume for the biomarker a random intercept and slope:
\begin{equation}
 Y_i(t)= m_i(t)+ \epsilon_i(t)={\boldsymbol X_{Li}(t)}^\top \boldsymbol \beta_L + b_{i0} + b_{i1}\times t +  \epsilon_i(t)\nonumber
\end{equation}
and then $b_{i0}$ represents the  heterogeneity of the baseline measures of the biomarker among the subjects and $b_{i1}$ the heterogeneity of the slope of the biomarker's linear trajectory among the subjects. 
As for Model (\ref{srem}) we allow the biomarker to be left-censored, again with $\boldsymbol y_i^o$ the observed outcomes and $\boldsymbol y_i^c$ the undetected ones. For individual $i$ we observe then $\boldsymbol \Theta_i=\{\boldsymbol y_i^o,\boldsymbol y_i^c,\boldsymbol T_i^{(1)},T_i,\boldsymbol \delta_i^{(1)},\delta_i\}$ and the interest is to estimate $\boldsymbol \xi=\{r_0(\cdot),\lambda_0(\cdot),\boldsymbol\beta_L,\boldsymbol\beta_R,\boldsymbol\beta_T,\boldsymbol B_1,\sigma_{v}^2,\sigma_{\epsilon}^2, \alpha,\boldsymbol\eta_R, \boldsymbol\eta_T\}$.

\subsection{Estimation}
\subsubsection{Maximum likelihood estimation}
Estimation of a model's parameters $\boldsymbol \xi$ is based on the maximization of the marginal log-likelihood derived from the joint distribution of the observed outcomes that are assumed to be independent from each other given the random effects. Let $\boldsymbol u_i$ represent random effects of a joint model ($\boldsymbol u_i$ can be a vector, for Models (\ref{jointgeneral}), (\ref{joint2recsurv}), (\ref{srem}), (\ref{trivariate})  or a scalar, for Model (\ref{jointrecsurv})) and $f_{\boldsymbol u_i}(\boldsymbol u_i;\boldsymbol \xi)$ the density function of the distribution of $\boldsymbol u_i$. The marginal individual likelihood is integrated over the random effects and is given by:
\begin{align}
\label{lik}
L_i(\boldsymbol \Theta_i;\boldsymbol \xi)=&\int_{\boldsymbol u_i}{f_{\boldsymbol y_i|\boldsymbol u_i}(\boldsymbol y_i|\boldsymbol u_i;\boldsymbol \xi)}^{\gamma_L}{f_{T^{(1)}_i|\boldsymbol u_i}(\boldsymbol T_i^{(1)},\delta_i^{(1)}|\boldsymbol u_i;\boldsymbol \xi)}^{\gamma_{R^{(1)}}}{f_{T^{(2)}_i|\boldsymbol u_i}(\boldsymbol T_i^{(2)},\delta_i^{(2)}|\boldsymbol u_i;\boldsymbol \xi)}^{\gamma_{R^{(2)}}}\notag\\
&\times f_{T_i|\boldsymbol u_i}(T_i,\delta_i|\boldsymbol u_i;\boldsymbol \xi)f_{\boldsymbol u_i}(\boldsymbol u_i;\boldsymbol \xi)\ d\boldsymbol u_i,
\end{align}
where indicators $\gamma_{\cdot}$ are introduced so that the likelihood is valid for all the joint models (1-5). Therefore, $\gamma_L=1$ in case of Models (\ref{srem}) and (\ref{trivariate}) and 0 otherwise, $\gamma_{R^{(1)}}=1$ in case of the models (\ref{jointrecsurv}), (\ref{joint2recsurv}) and (\ref{trivariate}) and 0 otherwise, $\gamma_{R^{(2)}}=1$ in case of the model (\ref{joint2recsurv}) and 0 otherwise.
 The conditional density of the longitudinal outcome $f_{\boldsymbol y_i|\boldsymbol u_i}$ is the density of the $l_i$-dimensional normal distribution with mean $\boldsymbol m_i=\{m_i(t_{ik}),k=1,\ldots,l_i\}$ and variance $\sigma_{\epsilon}^2\boldsymbol I_{m_i}$. In case of the left-censored data, we observe $l_i^o$ outcomes $\boldsymbol y_i^o$ and $l_i^c$ outcomes $\boldsymbol y_i^c$ are left-censored (below the threshold $s$). Then $f_{\boldsymbol y_i|\boldsymbol u_i}$ can be written as a product of the density for the observed outcomes (normal with mean $m_i(t_{ik})$ and variance $\sigma_{\epsilon}^2$) and the corresponding cumulative distribution function (cdf) $F_{\boldsymbol y_i|\boldsymbol u_i}$ for the censored outcomes:
\begin{align*}
f_{\boldsymbol y_i|\boldsymbol u_i}(\boldsymbol y_i|\boldsymbol u_i;\boldsymbol\xi)=\prod_{k=1}^{l_i^o}f_{y^o_{i}(t_{ik})|\boldsymbol u_i}(y^o_i(t_{ik})|\boldsymbol u_i;\boldsymbol\xi)\prod_{k=1}^{l_i^c}F_{y^c_{i}(t_{ik})|\boldsymbol u_i}(s|\boldsymbol u_i).
\end{align*}
Modeling of the risk of an event can be performed either in a calendar timescale or in a gap timescale. This choice depends on the interest of application whether the focus is on time from a fixed date, eg. beginning of a study, date of birth (calendar timescale) or on time between two consecutive events (gap timescale). The calendar timescale is often preferred in analyses of  disease evolution, for instance, in a randomized clinical trial. In this approach, the time of entering the interval for the risk of experiencing $j$th event is equal to the time of the $(j-1)$th event, an individual can be at risk of having the $j$th event only after the time of the $(j-1)$th event (delayed entry). On the other hand, in the gap timescale, after every recurrent event, the time of origin for the consecutive event is renewed to 0. In the medical context, this timescale is less natural than the calendar timescale, but it might be considered if the number of events is low or if the occurrence of the events does not affect importantly  subject's condition.

For the recurrent part of the model, the individual contribution from the recurrent event process of type $l$ ($l=1,2$) is given by the contribution from the right-censored observations and the event times and in the calendar timescale it is:
\begin{align*}
f_{T^{(l)}_i|\boldsymbol u_i}(\boldsymbol T_i^{(l)},\boldsymbol \delta_i^{(l)}|\boldsymbol u_i;\boldsymbol\xi)=\prod_{j=1}^{n_i^{(l)}}{\left(r_{ij}^{(l)}(T_{ij}^{(l)}|\boldsymbol u_i;\boldsymbol\xi)\right)}^{\delta_{ij}^{(l)}}\exp\left(-\int_{T_{i(j-1)}^{(l)}}^{T_{ij}^{(l)}}r_{ij}^{(l)}(t|\boldsymbol u_i;\boldsymbol\xi)dt\right).
\end{align*}
For the gap timescale the lower limit of the integral is 0 and the upper limit is the gap between the time of the $(j-1)$th event and the $j$th event \citep{Duchateau2003}. In case of only one type of recurrent events, $r_{ij}^{(l)}$ is obviously $r_{ij}$. Similarly, the individual contribution from the terminal event process is:
\begin{align*}
f_{T_i|\boldsymbol u_i}( T_i,\delta_i|\boldsymbol u_i;\boldsymbol\xi)={\left(\lambda_{i}(T_{i}|\boldsymbol u_i;\boldsymbol\xi)\right)}^{\delta_{i}}\exp\left(-\int_0^{T_{i}}\lambda_{i}(t|\boldsymbol u_i;\boldsymbol\xi)dt\right).
\end{align*}
For all the analyzed joint models the marginal likelihood (Equation~\ref{lik}) does not have an analytic form and integration is performed using quadrature methods. If a model includes one random effect that follows the Gamma distribution, the Gauss-Laguerre quadrature is used for the integral. The integrals over normally distributed random effects can be approximated using the Gauss-Hermite quadrature. 

For the multivariate joint frailty model (3), bivariate joint model for longitudinal data and a terminal event (4) and trivariate joint model (5) it is required (except of the bivariate joint model with a random intercept only) to approximate multidimensional integrals. In this case, the standard non-adaptive Gauss-Hermite quadrature, that uses a specific number of points, gives accurate results but often can be time consuming and thus alternatives have been proposed. The multivariate non-adaptive procedure using fully symmetric interpolatory integration rules proposed by \citet{Genz1996} offers advantageous computational time but in case of datasets in which some individuals have few repeated measurements, this method may be moderately unstable. Another possibility is the pseudo-adaptive Gauss-Hermite quadrature that uses transformed quadrature points to center and scale the integrand by utilizing estimates of the random effects from an appropriate linear mixed-effects model (this transformation does not include the frailty in the trivariate model, for which the standard method is used). This method enables using less quadrature points while preserving the estimation accuracy and thus lead to a better computational time \citep{Rizopoulos2012}.

\subsubsection{Parametric approach}

We propose to estimate the hazard functions using cubic M-splines, piecewise constant functions (PCF) or the Weibull distribution. In case of PCF and the Weibull distribution applied to the baseline hazard functions, the full likelihood is used for the estimation. Using the PCF for the baseline hazard function $h_0(t)$ (of recurrent events or a terminal event), an interval $[0,\tau]$ (with $\tau$ the last observed time among $N$ individuals) is divided into $n_{\textrm{int}}$ subintervals in one of two manners: equidistant so that all the subintervals are of the same length or percentile so that in each subinterval the same number of events is observed. Therefore, the baseline hazard function is expressed by
$
h_0(t)=\sum_{i=1}^{n_{\textrm{int}}}I_{\{t\in(t_{i-1},t_i)\}}c_i,
$
with $c_i\geq 0$. In the approach of PCF the crucial point is to choose the appropriate number of the intervals so that the estimated hazard function could capture enough the flexibility of the true function.
The other possibility in the parametric approach is to assume that the baseline hazard function $h_0(t)$ comes from the Weibull distribution with the shape parameter $a>0$ and the scale parameter $b>0$. Then the baseline hazard function is defined by $h_0(t)=(a t^{a-1})/b^{a}$. This approach is convenient given the small number of parameters to estimate (only two for each hazard function) but the resulting estimated functions are monotone and this constraint, in some cases, might be too limiting.

\subsubsection{Semi-parametric approach based using penalized likelihood}
In the semi-parametric approach, with regard to expected smooth baseline hazard functions, the likelihood of the model is penalized by terms depending on the roughness of the functions \citep{Joly1998}. 
 Cubic M-splines, polynomial functions of $3$th order are combined linearly to approximate the baseline hazard functions \citep{Ramsay1988}. 
	
For the estimated parameters $\boldsymbol \xi$ the full log-likelihood, $l_{joint}(\boldsymbol \xi)=\sum_{i=1}^N \textrm{ln}L_i(\boldsymbol \Theta_i,\boldsymbol \xi)$ is penalized in the following way:
\begin{equation}
 \begin{array}{lc}
pl(\boldsymbol \xi)=l_{joint}(\boldsymbol \xi)-\kappa_1\int_{0}^{\infty}{r_0^{''}(t)^{2}dt}-\kappa_2\int_{0}^{\infty}{\lambda_0^{''}(t)^{2}dt}, & \textrm{for Model (\ref{jointrecsurv}),(\ref{jointgeneral}),(\ref{trivariate})} \\
pl(\boldsymbol \xi)=l_{joint}(\boldsymbol \xi)-\kappa_1\int_{0}^{\infty}{r_0^{(1)''}(t)^{2}dt}-\kappa_2\int_{0}^{\infty}{r_0^{(2)''}(t)^{2}dt}-\kappa_3\int_{0}^{\infty}{\lambda_0^{''}(t)^{2}dt},& \textrm{for Model (\ref{joint2recsurv})} \\
pl(\boldsymbol \xi)=l_{joint}(\boldsymbol \xi)-\kappa_1\int_{0}^{\infty}{\lambda_0^{''}(t)^{2}dt},& \textrm{for Model (\ref{srem})} \nonumber\\
\end{array},
\end{equation}
The positive smoothing parameters ($\kappa_1$, $\kappa_2$ and $\kappa_3$) provide a balance between the data fit and the smoothness of the functions. Both the full and penalized log-likelihood are maximized using the robust Marquardt algorithm \citep{Marquardt1963}, a mixture between the Newton-Raphson and the steepest descent algorithm.

\subsubsection{Goodness of fit}
For verification of models assumptions, residuals are used as a standard statistical tool for the graphical assessment. In the context of the survival data (recurrent and terminal events) these are the martingale residuals and in the context of the longitudinal data, both,  the residuals conditional on random effects and the marginal residuals are often used.

The martingale residuals  model whether the number of observed events is correctly predicted by a model. Principally, it is based on the counting processes theory and for  subject $i$ and time $t$ they are defined as the difference between the number of events of  subject $i$ until $t$ and the Breslow estimator of the cumulative hazard function of $t$. Let $N_i(t)$ be the counting process of the event of type $p$ (recurrent or terminal), $\boldsymbol u_i$ represent the random effects and the process's intensity $\zeta_i^{(p)}(t|\boldsymbol u_i)=\boldsymbol u_i\zeta_0^{(p)}(t)\exp(\boldsymbol X_{pi}(t)^\top \boldsymbol \beta_p)=\boldsymbol u_i\zeta^{(p)}_i(t)$ with $\zeta_0^{(p)}(t)$ the baseline risk and $\boldsymbol X_{pi}(t)$ the prognostic factors. The martingale residuals can be expressed by (for details, see \citet{Commenges2000}):
\begin{equation}
M_i(t)=N_i(t)-\widehat{\boldsymbol u_i}\int_0^t W_i(s)\widehat{\zeta_i^{(p)}}(s) ds,
\end{equation}
where $W_i(t)$ is equal to 1 if the individual is at risk of the event at time $t$ and 0 otherwise. The martingale residuals are calculated at $T_i$, that is at the end of the follow-up (terminal event). The assessment of the model is performed visually, the mean of the martingale residuals at a given time point should be equal to 0.

For the longitudinal data estimated in the framework of the linear mixed effects models, there exist marginal residuals averaged on the population level defined by $\boldsymbol R_i^{(m)}=\boldsymbol y_i-\boldsymbol X_{Li}^\top \hat{\boldsymbol \beta}_L$ and conditional residuals that are subject-specific, $\boldsymbol R_i^{(c)}=\boldsymbol y_i-\boldsymbol X_{Li}^\top \hat{\boldsymbol \beta}_L-\boldsymbol Z_i^\top \hat{\boldsymbol b}_i$. These raw residuals are recommended for checking homoscedasticity  of the conditional and marginal variance. For verification of the normality assumption and detection of outlying observations, the Cholesky residuals are more adapted as they represent decorrelated residuals and are defined by:
\begin{equation}
\boldsymbol R_i^{(m)*}=\widehat{\boldsymbol U^{(m)}_i}\boldsymbol R_i^{(m)},\hspace{0.5cm} \boldsymbol R_i^{(c)*}=\widehat{\boldsymbol U^{(c)}_i}\boldsymbol R_i^{(c)} \nonumber
\end{equation}
where the raw residuals are multiplied by the upper-triangular matrices ($\widehat{\boldsymbol U^{(m)}_i}$ and $\widehat{\boldsymbol U^{(c)}_i}$) obtained by the Cholesky decomposition of the variance-covariance matrices:
\begin{equation}
\begin{array}{c}
\boldsymbol V_{\boldsymbol R_i^{(m)}}=\widehat{\boldsymbol V_i}-\boldsymbol X_{Li}(\sum_{i=1}^{N}\boldsymbol X_{Li}^\top \widehat{\boldsymbol V_i}^{-1}\boldsymbol X_{Li})^{-1}\boldsymbol X_{Li}^\top={\widehat{\boldsymbol U^{(m)}_i}}^\top\widehat{\boldsymbol U^{(m)}_i}, \\ \boldsymbol V_{\boldsymbol R_i^{(c)}}=\widehat{\sigma_{\epsilon}}^2\boldsymbol I_{n_i}\widehat{\boldsymbol V_i}^{-1}\boldsymbol V_{\boldsymbol R_i^{(m)}}\widehat{\boldsymbol V_i}^{-1}
\widehat{\sigma_{\epsilon}}^2\boldsymbol I_{n_i}={\widehat{\boldsymbol U^{(c)}_i}}^\top\widehat{\boldsymbol U^{(c)}_i}
\nonumber,
\end{array}
 \end{equation}
where $\boldsymbol V_i$ is the marginal variance-covariance matrix of the longitudinal outcome $\boldsymbol y_i$, equal to $\boldsymbol Z_i\textbf{B}_1\boldsymbol Z_i^\top+\sigma_{\epsilon}^2 \boldsymbol I_{n_i}$ and of dimension $l_i\times l_i$. The marginal Cholesky residuals should be approximately normally distributed and thus, their normal Q-Q plot allows to verify the assumption of the normality.

Both, for the calculation of the martingale residuals and the residuals of the longitudinal outcome, the values of the random effects are necessary. For this purpose, the empirical Bayes estimators of $\boldsymbol u_i$  are calculated using the formula for the posterior probability function:
\begin{equation}
f(\boldsymbol u_i|\boldsymbol \Theta_i;\widehat{\boldsymbol \xi})=\frac{f(\boldsymbol \Theta_i|\boldsymbol u_i;\widehat{\boldsymbol \xi})f(\boldsymbol u_i;\widehat{\boldsymbol \xi})}{f(\boldsymbol \Theta_i;\widehat{\boldsymbol \xi})}\propto f(\boldsymbol \Theta_i|\boldsymbol u_i;\widehat{\boldsymbol \xi})f(\boldsymbol u_i;\widehat{\boldsymbol \xi}) .\nonumber\end{equation}
For the joint models, this expression does not have an analytical solution and the numerical computation is applied that finds such $\boldsymbol u_i$ that maximizes $f(\boldsymbol u_i|\boldsymbol \Theta_i;\widehat{\boldsymbol \xi})$:
\begin{equation}
\widehat{\boldsymbol u_i}=\argmax_{\boldsymbol u_i} f(\boldsymbol u_i|\boldsymbol  \Theta_i;\widehat{\boldsymbol \xi}),\nonumber
\end{equation}
and this is obtained using the Marquardt algorithm.

\section{Prediction of risk of the terminal event}
\label{sec:prediction}
\begin{figure}
	\includegraphics[ trim = 0.0cm 5.5cm 1.5cm 2cm, clip]{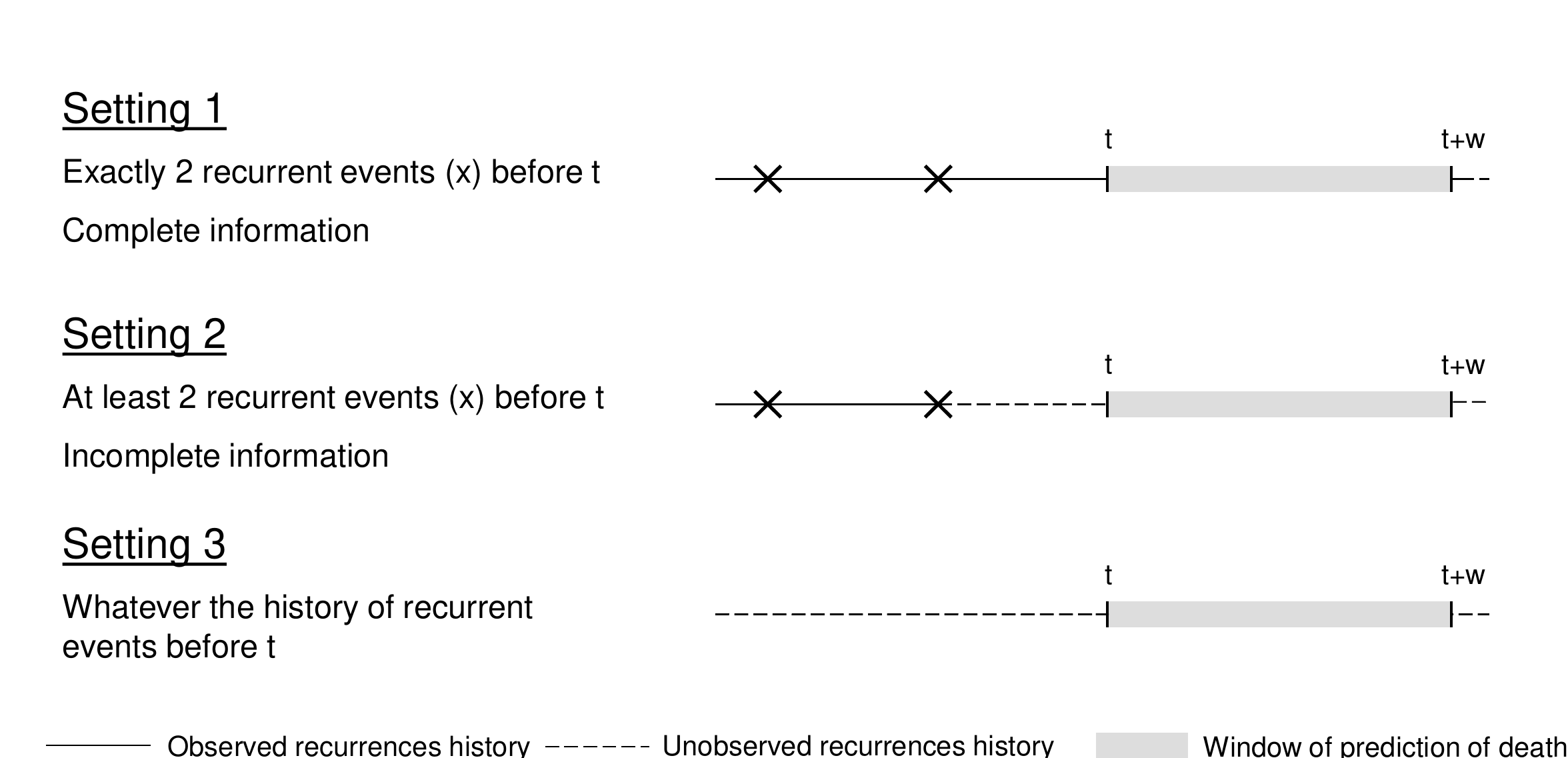} 
	\caption{Three settings to take into account patient history of recurrent events in prediction. Illustration with two recurrent events.}
	\label{figure:pred:settings}
\end{figure}

Specific predictions can be obtained in the framework of joint modeling. Prediction consists of estimating the probability of an event at a given time $t+w$ knowing the available information at prediction time $t$. Using a joint model, it is possible to estimate the probability of having the terminal event at time $t+w$ given the history of the individual (occurrences of the recurrent events and/or the measurements of the longitudinal biomarker) prior to $t$. For the joint model for recurrent events and a terminal event (\ref{jointrecsurv}), three settings of prediction were developed by \citet{Mauguen2013}. In the first one, all the available information is accounted for, and we consider that this information is complete. In the second setting, all the known information is accounted for, however we consider that this information may be incomplete. Finally, in the third setting, recurrence information is not accounted for and only covariates are considered. All the settings are represented in Figure~\ref{figure:pred:settings}. Here, we focus on the first setting, for which we present the predictions and the accurate measures of predictive accuracy but in the package all the three setting are implemented for the joint models for recurrent and terminal events.

\begin{figure}
	\includegraphics[ trim = 0.2cm 3.5cm 0.4cm 2.2cm, clip]{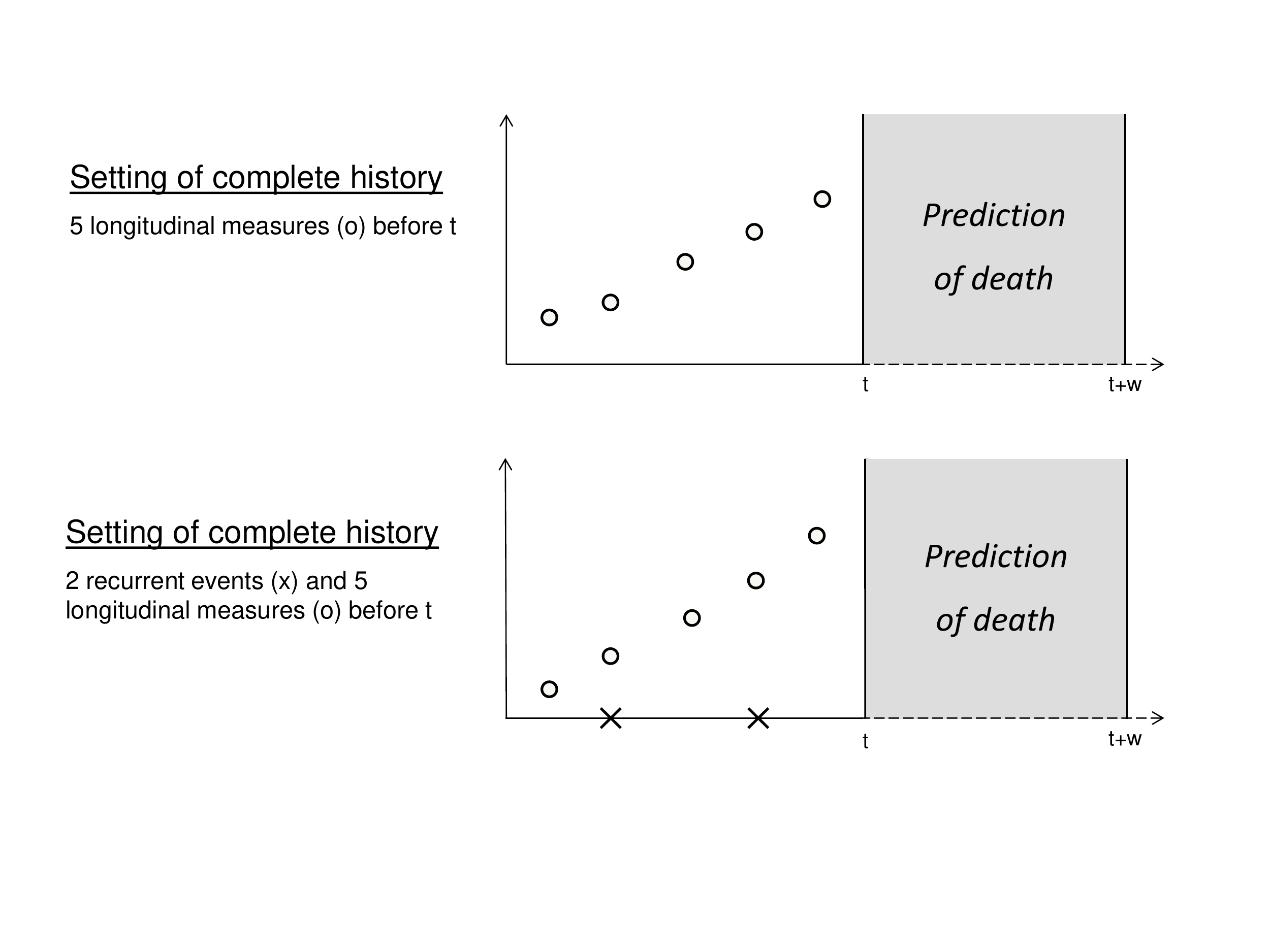} 
	\caption{The possible prediction settings including the longitudinal data and considering the whole information available. The top setting correspond to the Model (\ref{srem}) and the bottom graphic to the Model (\ref{trivariate}).}
	\label{figure:pred:longi}
\end{figure}

 For the joint models with a longitudinal outcome a complete history of the biomarker is considered \citep{Krol2015}. Thus, for Model (\ref{srem}) the individual's history is the whole observed trajectory of the biomarker and for Model (\ref{trivariate}) it is the the whole observed trajectory of the biomarker and all the observed occurrences of the recurrent event (see Figure~\ref{figure:pred:longi}).

The proposed prediction can be performed for patients from the population used to develop the model, but also for \lq\lq{}new patients\rq\rq{} from other populations. This is possible as the probabilities calculated are marginal, i.e., the conditional probabilities are integrated over the distribution of the random effects. Thus, values of patients frailties are not required to estimate the probabilities of the event. However, it should be noted that the predictions include individual deviations via the estimated parameters of the random effects' distribution in a joint model.

We denote by $t$ the time at which the prediction is made and by $w$ the window of prediction. Thus, we are interested in the probability of the event (death) at time $t+w$, knowing what happened before time $t$. The general formulation of predicted probability of the terminal event conditional on random effects and patient's history is:
\begin{equation}
 \begin{array}{ll}
\label{prediction}
P_i(t,t+w;\boldsymbol \xi)& = P(T_i^*\le t+w|T_i^*> t, \mathcal{F}_{i}(t),\boldsymbol  X_{i}; \boldsymbol \xi) \\
\end{array}
\end{equation}
where $\boldsymbol  X_{i}$ are all the covariates included in the model, $\mathcal{F}_{i}(t)$ corresponds to the complete repeated data of patient $i$ observed until time $t$. We define the complete history of recurrences $\mathcal{H}_{i}^{J(l)}(t)=\{N_i^{R(l)}(t)=J^{(l)},  T_{i1}^{(l)*}<\cdots<T_{iJ}^{(l)*} \le t \}$ ($N_i^{R}(t)$ is the counting process of the recurrent events, $l=1,2$ for the two types of recurrent events, and in case of only one type of recurrent events in the models the index $(l)$ is omitted) and the history of the biomarker  $\mathcal{Y}_i(t)=\{y_i(t_{iK}),t_{i1}<\cdots<t_{iK}<t\}$. Therefore, the individual's history $\mathcal{F}_{i}(t)$ depends on the model considered and is equal to: 
\begin{equation}
 \begin{array}{lc}
\mathcal{F}_{i}(t)=\mathcal{H}_{i}^{J}(t),& \textrm{for Model (\ref{jointrecsurv}),   (\ref{jointgeneral})} \\
\mathcal{F}_{i}(t)=\{\mathcal{H}_{i}^{J(1)}(t),\mathcal{H}_{i}^{J(2)}(t)\}, &  \textrm{for Model (\ref{joint2recsurv})} \\ \nonumber
\mathcal{F}_{i}(t)=\mathcal{Y}_i(t), & \textrm{for Model (\ref{srem})} \\ \nonumber
\mathcal{F}_{i}(t)=\{\mathcal{H}_{i}^{J}(t),\mathcal{Y}_i(t)\},  &\textrm{for Model (\ref{trivariate})}  \nonumber
\end{array}
\end{equation}
and for the recurrent events we assume $T_{i0}^*=0$ and $T_{i(J+1)}^*>t$. For the estimated probabilities, confidence intervals are obtained by the Monte Carlo (MC) method, using the $2.5^{th}$ and $97.5^{th}$ percentiles of the obtained distribution (percentile confidence interval).

\subsection{Brier score}
In order to validate the prediction ability of a given model, a prediction error is proposed using the weighted Brier score \citep{Gerds2006,Mauguen2013}. It measures the distance between the prediction (probability of event) and the actual outcome (dead or alive). Inverse probability weighting corrects the score for the presence of right censoring. At a given horizon of prediction $t+w$, the error of prediction is calculated by:
\begin{equation}
Err_{t+w} = 	\frac{1}{N_t} \sum\limits_{i=1}^{N_t}  [ I(T^*_i>t+w)-(1- \widehat{P}_i(t,t+w;\widehat{\boldsymbol \xi})) ]^2 \widehat{w_i}(t+w,\widehat{G_N})
\end{equation}
	where the weights $\widehat{w_i}(t+w,\widehat{G_n})$ is defined by:
	\begin{equation}
	\widehat{w_i}(t+w,\widehat{G_n}) = \frac{I(T^*_i \leq t+w)\delta_i}{\widehat{G_N}(T^*_i)/\widehat{G_N}(t)} + \frac{I(T^*_i > t+w )}{\widehat{G_N}(t+w)/\widehat{G_N}(t)},\nonumber 
	\end{equation}
and $\widehat{G_N}(t)$ is the Kaplan-Meier estimate of the population censoring distribution at time $t$, $N_t$ is the number of patients at risk of the event (alive and uncensored).

Direct calculations of the Brier score are not implemented in the package \pkg{frailtypack} but using predictions $\widehat{P}_i(t,t+w;\widehat{\boldsymbol \xi}))$, the Brier score can be obtained using weights $\widehat{w_i}(\cdot)$ function \code{pec} from the package \pkg{pec} \citep{pec} (for details, see the colorectal example in Section~\ref{colorectal}). For an internal validation (on a training dataset, i.e., used for estimation) of the model as a prediction tool for new patients, a $k$-fold cross-validation is used to correct for over-optimism. In this procedure, the joint model estimations are performed $k$ times on $k-1$ partitions from the random split and the predictions are calculated on the left partitions. The prediction curves are based on these predictions, in the calculated for all the patients.

\subsection{EPOCE}
Another method to evaluate a model's predictive accuracy is the EPOCE estimator that is derived using prognostic conditional log-likelihood \citep{Commenges2012}. This measure is adapted both for an external data and then the Mean Prognosis Observed Loss (MPOL) is computed, as well as for the training data using the approximated Cross-Validated Prognosis Observed Loss (CVPOL$_a$). 

This measure of predictive accuracy is the risk function of an estimator of the true prognostic density function $f^*_{T|\mathcal{F}(t),T^*\geq t}$, where $\mathcal{F}(t)$ denotes the history of repeated measurements and/or recurrent events until time $t$. Using information theory this risk is defined as the expectation of the loss function, that is the estimator derived from the joint model $f_{T|\mathcal{F}(t),T^*\geq t}$ conditioned on $T^*>t$ and this can be written as $\textrm{EPOCE}(t)=\mathbb{E}(-\textrm{ln}(f_{T|\mathcal{F}(t),T^*>t})|T^*>t)$. In case of the model evaluated on the training data the approximated leave-one-out CVPOL$_a$ is defined by:
\begin{equation}
\label{CVPOL}
\textrm{CVPOL}_a(t)=-\frac{1}{N_t}\sum_{i=1}^{N_t}F_i(\hat{\boldsymbol \xi}_i,t)+N \textrm{Trace}(\boldsymbol H^{-1}\boldsymbol K_t),
\end{equation}
where  $\boldsymbol H^{-1}$ is the inverted hessian matrix of the joint log-likelihood (Equation~\ref{lik}), $\boldsymbol K_t=\frac{1}{N_t(N-1)}\sum_{i=1}^{N}I_{\{T_i>t\}}\hat{\boldsymbol v}_i \hat{\boldsymbol d}_i^{\top}$ with $\hat{\boldsymbol v}_i=\frac{\partial F_i(\boldsymbol \xi,t)}{\partial \boldsymbol \xi}|_{\hat {\boldsymbol \xi}} $ and  $\hat{\boldsymbol d}_i=\frac{\partial l_i(\boldsymbol \Theta,\boldsymbol \xi)}{\partial \boldsymbol \xi}|_{\hat{\boldsymbol \xi}}$. The individual contribution  to the log-likelihood  of a terminal event at $t$ defined for the individuals that are still at risk of the event at $t$ can be written as:
\begin{equation}
F_i(\hat{\boldsymbol \xi},t)=
\textrm{ln}\left(\frac{\int_{\boldsymbol u_i}f_{\mathcal{Y}_i(t)|\boldsymbol u_i}(\mathcal{Y}_i(t)|\boldsymbol u_i;\hat{\boldsymbol \xi})^{\gamma_L}\displaystyle\prod_{k=1}^l f_{\mathcal{H}^{J(k)}_i(t)|\boldsymbol u_i}(\mathcal{H}^{J(k)}_i(t)|\boldsymbol u_i;\hat{\boldsymbol \xi})^{\gamma_R^{(k)}}f_{T_i|\boldsymbol u_i}(T_i,\delta_i|\boldsymbol u_i;\hat{\boldsymbol \xi})f_{\boldsymbol u_i}(\boldsymbol u_i)\ d\boldsymbol u_i}{\int_{\boldsymbol u_i}f_{\mathcal{Y}_i(t)|\boldsymbol u_i}(\mathcal{Y}_i(t)|\boldsymbol u_i;\hat{\boldsymbol \xi})^{\gamma_L}\displaystyle\prod_{k=1}^l f_{\mathcal{H}^{J(k)}_i(t)|\boldsymbol u_i}(\mathcal{H}^{J(k)}_i(t)|\boldsymbol u_i;\hat{\boldsymbol \xi})^{\gamma_R^{(k)}}S_i^t(T_i|\boldsymbol u_i;\hat{\boldsymbol \xi})f_{\boldsymbol u_i}(\boldsymbol u_i)\ d\boldsymbol u_i} \right),\nonumber
\end{equation}
where $S_i^t$ is the survival function of the terminal event for individual $i$. Again, $l$ denotes the number of types of recurrent events included in the model.

In case of the external data the $\textrm{MPOL}(t)$ is expressed by the first component of the sum in the CVPOL$_a$(t) (Equation~\ref{CVPOL}). Indeed, the second component corresponds to the statistical risk introduced to the CVPOL$_a$ in order to correct for over-optimism coming from the use of the approximated cross-validation. The comparison of the models using the EPOCE can be done visually by tracking the 95\% intervals of difference in EPOCE. 

The EPOCE estimator has some advantages comparing to the prediction error, in particular for the training data the approximated cross-validation technique is less computationally demanding than the crude leave-one-out cross-validation method for the prediction error, which is important in case of joint models when time of calculation is often long \citep{Proust-Lima2014}.

\section{Modeling and prediction using the R package frailtypack}
\label{package}
\subsection{Estimation of joint models}
In the package \pkg{frailtypack} there are four different functions for the estimation of joint models, one for each model. The joint models for recurrent events and a terminal event (\ref{jointrecsurv}) (and models for two clustered survival outcomes) as well as the general joint frailty models (\ref{jointgeneral})  are estimated with function \code{frailtyPenal}. The joint models for two recurrent events and a terminal event (\ref{joint2recsurv}) are estimated with \code{multivPenal}. The estimation of joint models for a longitudinal outcome and a terminal event (\ref{srem}) is performed with \code{longiPenal}. Finally, function \code{trivPenal} estimates trivariate joint models (\ref{trivariate}). All these functions make calls to  compiled \proglang{Fortran} codes programmed for computation and optimization of the log-likelihood. In the following, we detail each of the joint models functions.

\vspace{0.2cm}
\code{frailtyPenal} \textit{function}
\begin{CodeChunk}
\begin{CodeInput}
frailtyPenal(formula, formula.terminalEvent, data, recurrentAG = FALSE,
  cross.validation = FALSE, jointGeneral, n.knots, kappa, maxit = 300, 
  hazard = "Splines", nb.int, RandDist = "Gamma", betaknots = 1, 
  betaorder = 3, initialize = TRUE, init.B, init.Theta, init.Alpha, Alpha, 
  init.Ksi, Ksi, init.Eta, LIMparam = 1e-3, LIMlogl = 1e-3, LIMderiv = 1e-3, 
  print.times = TRUE)
\end{CodeInput}
\end{CodeChunk}
Argument \code{formula} is a two-sided formula for a survival object \code{Surv} from the \pkg{survival} package \citep{survival} and it represents the recurrent event process (the first survival outcome for the joint models in case of clustered data) with the combination of covariates on the right-hand side, the indication of grouping variable (with term \code{cluster(group)}) and the indication of the variable for the terminal event (e.g., \code{terminal(death)}). It should be noted that the function \code{cluster(x)} is different from that included in the package \pkg{survival}. In both cases it is used for the identification of the correlated groups but in \pkg{frailtypack} it indicates the application of frailty model and in \pkg{survival}, a GEE (Generalized Estimating Equations) approach, without random effects. Argument \code{formula.terminalEvent} requires the combination of covariates related to the terminal event  on the right-hand side. The name of  \code{data.frame} with the variables used in the function should be put for argument \code{data}. Logical argument \code{recurrentAG} indicates whether the calendar timescale for recurrent events or clustered data with the counting process approach of \citet{Andersen1982} (\code{TRUE}) or the gap timescale (\code{FALSE} by default) is to be used. Argument for the cross-validation \code{cross.validation} is not implemented for the joint models, thus its logical value must be \code{FALSE}. The smoothing parameters $\kappa$ for a joint model can be chosen by first fitting suitable shared frailty and Cox models with the cross-validation method.

The general joint frailty models (\ref{jointgeneral}) can be estimated if argument \code{jointGeneral} is \code{TRUE}. In this case, the additional gamma frailty term is assumed and the $\alpha$ parameter is not considered. These models can be applied only with the Gamma distribution for the random effects. 

The type of approximation of the baseline hazard functions is defined by argument \code{hazard} and can be chosen among \code{Splines} for semiparametric functions using equidistant intervals, \code{Splines-per} using percentile intervals, \code{Piecewise-equi} and \code{Piecewise-per} for piecewise constant functions using equidistant and percentile intervals, respectively and \code{Weibull} for the parametric Weibull baseline hazard functions. If \code{Splines} or \code{Splines-per} is chosen for the baseline hazard functions,  arguments \code{kappa} for the positive smoothing parameters and \code{n.knots} should be given with the number of knots chosen between 4 and 20 which corresponds to \code{n.knots}+2 splines functions for approximation of the baseline hazard functions (the same number for hazard functions for both outcomes). If \code{Percentile-equi} or \code{Percentile-per} is chosen for the approximation, argument \code{nb.int} should be given with a 2-element vector of numbers of time intervals (1-20) for the two baseline hazard functions of the model.

Argument \code{RandDist} represents the type of the random effect distribution, either \code{Gamma} for the Gamma distribution or \code{LogN} for the normal distribution (log-normal joint model). If it is assumed that $\alpha$ in Model (\ref{jointrecsurv}) is equal to zero, argument \code{Alpha} should be set to \code{"None"}.

In case of time dependent covariates, arguments \code{betaknots} and \code{betaorder} are used for the number of inner knots used for the estimation of B-splines (1, by default) and the order of B-splines (3 for quadratic B-splines, by default), respectively.

The rest of the arguments are allocated for the optimization algorithm. Argument \code{maxit} declares the maximum number of iterations for the Marquardt algorithm. For a joint nested frailty model, i.e., a model that allow joint analysis of recurrent and terminal events for hierarchically clustered data, argument \code{initialize} determines whether the parameters should be initialized with estimated values from the appropriate nested frailty models. Arguments \code{init.B}, \code{init.Theta}, \code{init.Eta} and \code{init.Alpha} are vectors of initial values for regression coefficients, variances of the random effects and for the $\alpha$ parameter (by default, 0.5 is set for all the parameters). Arguments \code{init.Ksi} and \code{Ksi} are defined for joint nested frailty models and correspond to initial value for the flexibility parameter and the logical value indicating whether include this parameter in the model or not, respectively. The convergence thresholds of the Marquardt algorithm are for the difference between two consecutive log-likelihoods (\code{LIMlogl}), for the difference between the consecutive values of estimated coefficients (\code{LIMparam}) and for the small gradient of the log-likelihood (\code{LIMderiv}). All these threshold values are $10^{-3}$ by default. Finally, argument \code{print.times} indicates whether to print the iteration process (the information note about the calculation process and time taken by the program after terminating), the default is \code{TRUE}. 

The function \code{frailtyPenal} returns objects of class \code{jointPenal} if joint models are estimated. It should be noted that using this function univariate models: shared frailty models \citep{Rondeau2012frailtypack} and Cox models can be applied as well resulting with objects of class \code{frailtyPenal}. For both classes methods \code{print()}, \code{summary()} and \code{plot()} are available.

\vspace{0.2cm}
\code{multivPenal} \textit{function}
\begin{CodeChunk}
\begin{CodeInput}
multivPenal(formula, formula.Event2, formula.terminalEvent, data,
  initialize = TRUE, recurrentAG = FALSE, n.knots, kappa, maxit = 350,
  hazard = "Splines", nb.int, print.times = TRUE)
\end{CodeInput}
\end{CodeChunk}
This function allows to fit the multivariate frailty models (\ref{joint2recsurv}). Argument \code{formula} must be a two-sided formula for a \code{Surv} object, corresponding to the first type of the recurrent event (no interval-censoring is allowed). Arguments \code{formula.Event2} refer to the second type of the recurrent event and \code{formula.terminalEvent} to the terminal event, and are equal to linear combinations related to the respective events. The rest of the arguments is analogical to \code{frailtyPenal}. Arguments \code{n.knots} (values between 4 and 20) and \code{kappa} must be vectors of length 3 for each type of event, first for the recurrent event of type 1, second for the recurrent event of type 2 and third for the terminal event. The function \code{multivPenal} return objects of class \code{multivPenal} with \code{print()}, \code{summary()} and \code{plot()} methods available.

\vspace{0.2cm}
\texttt{longiPenal} \textit{function}
\begin{CodeChunk}
\begin{CodeInput}
longiPenal(formula, formula.LongitudinalData, data, data.Longi, random, id, 
  intercept = TRUE, link = "Random-effects", left.censoring = FALSE, n.knots,
  kappa, maxit = 350, hazard = "Splines", nb.int, init.B, init.Random, 
  init.Eta, method.GH = "Standard", n.nodes, LIMparam = 1e-3, LIMlogl = 1e-3,
  LIMderiv = 1e-3, print.times = TRUE)
\end{CodeInput}
\end{CodeChunk}
In this function for the joint analysis of a terminal event and a longitudinal outcome, argument \code{formula} refers to the terminal event and the left-hand side of the formula must be a \code{Surv} object and the right-hand side, the explicative variables. Argument \code{formula.LongitudinalData} is equal to the sum of fixed explicative variables for the longitudinal outcome. For the model, two datasets are  required: \code{data} containing information on the terminal event process and \code{data.Longi} with data related to longitudinal measurements. 
Random effects associated to the longitudinal outcome are defined with \code{random} using the appropriate names of the variables from  \code{data.Longi}. If a random intercept is assumed, the option \code{"1"} should be used. For a random intercept and slope, arguments \code{random} should be equal to a vector with elements \code{"1"} and the name of a variable representing time point of the biomarker measurements. At the moment, more complicated structures of the random effects are not available in the package. The name of the variable representing the individuals in  \code{data.Longi} is indicated by \code{id}. The logical argument \code{intercept} determines whether a fixed intercept should be included in the longitudinal part or not (default is \code{TRUE}). Two types of subject-specific link function are to choose and defined with the argument \code{link}. The default option \code{"Random-effects"} represents the link function of the random effects $\textbf{b}_i$, otherwise the option is  \code{"Current-level"} for the link function of the current level of the longitudinal outcome $m_i(t)$. 
 
The initial values of the estimated parameters can be indicated by \code{init.B} for the fixed covariates (a vector of values starting with the parameters associated to the terminal event and then for the longitudinal outcome, interactions in the end of each component), \code{init.Random} for the vector of elements of the Cholesky decomposition of the covariance matrix of the random effects $\textbf{u}_i$ and \code{init.Eta} for the regression coefficients associated to the link function.

There are three methods of the Gaussian quadrature to approximate the integrals to choose from. The default \code{Standard} corresponds to the non-adaptive Gauss-Hermite quadrature for multidimensional integrals. The other possibility is the pseudo-adaptive Gaussian quadrature (\code{"Pseudo-adaptive"}) \citep{Rizopoulos2012}. Finally, the multivariate non-adaptive Gaussian quadrature using the algorithm implemented in a \proglang{Fortran} subroutine HRMSYM is indicated by \code{"HRMSYM"} \citep{Genz1996}. The number of the quadrature nodes (\code{n.nodes}) can be chosen among 5, 7, 9, 12, 15, 20 and 32 using argument (default is 9). 

The function \code{longiPenal} return objects of class \code{longiPenal} with \code{print()}, \code{summary()} and \code{plot()} methods available.

\vspace{0.2cm}
\texttt{trivPenal} \textit{function}
\begin{CodeChunk}
\begin{CodeInput}
trivPenal(formula, formula.terminalEvent, formula.LongitudinalData, data, 
  data.Longi, random, id, intercept = TRUE, link = "Random-effects",
  left.censoring = FALSE, recurrentAG = FALSE, n.knots, kappa, maxit = 300,
  hazard = "Splines", nb.int, init.B, init.Random, init.Eta, init.Alpha, 
  method.GH = "Standard", n.nodes, LIMparam = 1e-3, LIMlogl = 1e-3, 
  LIMderiv = 1e-3, print.times = TRUE)
\end{CodeInput}
\end{CodeChunk}
The function for the trivariate joint model comprises three formulas for each type of process. The first two arguments are analogous to \code{frailtyPenal}, argument \code{formula}, referring to recurrent events, is a two-sided formula for a \code{Surv} object on the left-hand side and covariates on the right-hand side (with indication of the variable for the terminal event using method \code{terminal}) and argument \code{formula.terminalEvent} represents the terminal event and is equal to a linear combination of the explicative variables. Finally, argument \code{formula.LongitudinalData} as in function \code{longiPenal} corresponds to the longitudinal outcome indicating the fixed effect covariates.  The rest of the arguments are detailed in the descriptions of functions \code{frailtyPenal} and \code{longiPenal}. 
The function \code{trivPenal} return objects of class \code{trivPenal} with \code{print()}, \code{summary()} and \code{plot()} methods available.

\subsection{Prediction}
The current increase of interest in the joint modeling of correlated data is often related to the individual predictions that these models offer. Indeed, calculating the probabilities of a terminal event given a joint model results in precise predictions that consider the past of an individual. Moreover, there exist statistical tools that evaluate a model's capacity for these dynamic predictions. In the package \pkg{frailtypack} we provide \code{prediction} function for dynamic predictions of a terminal event in a finite horizon, \code{epoce} function for evaluating predictive accuracy of a joint model and \code{Diffepoce} for comparing the accuracy of two joint models.

\vspace{0.2cm}
\textit{Predicted probabilities with} \texttt{prediction} \textit{function}\\
In the package it is possible to calculate the prediction probabilities for the Cox proportional hazard, shared frailty (for clustered data, \citet{Rondeau2012frailtypack}) and joint models. Among the joint models the predictions are provided for the standard joint frailty models (recurrent events and a terminal event), for the joint models for a longitudinal outcome and a terminal event and for the trivariate joint models (a longitudinal outcome, recurrent events and a terminal event). These probabilities can be calculated for a given prediction time and a window or for a vector of times, with varying prediction time or varying window. 
For the shared frailty models for clustered events, marginal and conditional on a specific cluster predictions can be calculated and for the joint models only the marginal predictions are provided. Finally, for the joint frailty models the predictions are calculated in three settings: given the exact history of recurrences, given the partial history of recurrences (the first $J$ recurrences) and ignoring the past recurrences. For the joint models with a longitudinal outcome (bivariate and trivariate) only the predictions considering the patient's complete history are provided. For all the aforementioned predictions the following function is proposed:
\begin{CodeChunk}
\begin{CodeInput}
prediction(fit, data, data.Longi, t, window, group, MC.sample = 0)
\end{CodeInput}
\end{CodeChunk}
Argument \code{fit} indicates the, a \code{frailtyPenal}, \code{jointPenal}, \code{longiPenal} or \code{trivPenal} object. The data with individual characteristics for predictions must be provided in dataframe \code{data} with information on the recurrent events and covariates related to recurrences and the terminal event and in case of \code{longiPenal} and \code{trivPenal} dataframe \code{data.Longi} containing repeated measurements and covariates related to the longitudinal outcome. These two datasets must refer to the same individuals for which the predictions will be calculated. Moreover, the names and the types of variables should be equivalent to those in the dataset used for estimation. The details on how to prepare correctly the data are presented in appropriate examples (Section~\ref{examples}).

Argument \code{t} is a time or vector of times for predictions and \code{window} is a horizon or vector of horizons. The function calculates the probability of the terminal event between a time of prediction and a horizon (both arguments are scalars), between several times of prediction and a horizon (\code{t} is a vector and \code{window} a scalar) and finally, between a time of prediction and several horizons (\code{t} is a scalar and \code{window} a vector of positive values).

For all the predictions, confidence intervals can be calculated using the MC method with \code{MC.sample} number of samples (maximum 1000). If the confidence bands are not to be calculated  argument \code{MC.sample} should be equal to 0 (the default).

\vspace{0.2cm}
\textit{Predictive accuracy measure with} \texttt{epoce} \textit{and}  \texttt{Diffepoce} \textit{functions}\\
Predictive ability of joint models can be evaluated with function \code{epoce} that computes the estimators of the EPOCE, the MPOL and CVPOL$_a$. For a given estimation, the evaluation can be performed on the same data and then both MPOL and CVPOL$_a$ are calculated, as well as on a new dataset but then only MPOL is calculated as the correction for over-optimism is not necessary. The call of the function is:
\begin{CodeChunk}
\begin{CodeInput}
epoce(fit, pred.times, newdata = NULL, newdata.Longi = NULL)
\end{CodeInput}
\end{CodeChunk}
with \code{fit} an object of \code{jointPenal}, \code{longiPenal} or \code{trivPenal} classes, \code{pred.times} a vector of time for which the calculations are performed. In case of external validation, new datasets \code{newdata} and \code{newdata.Longi} should be provided  (\code{newdata.Longi} only in case of \code{longiPenal} and \code{trivPenal} objects). However, the names and types of variables in \code{newdata} and \code{newdata.Longi} must be the same as in the datasets used for estimation.

To compare the predictive accuracy of two joint models fit to the same data but possibly with different covariates, the simple comparison of obtained values of EPOCE can be enhanced by calculating the 95\% tracking interval of difference between the EPOCE estimators. For this purpose we propose function:
\begin{CodeChunk}
\begin{CodeInput}
Diffepoce(epoce1, epoce2)
\end{CodeInput}
\end{CodeChunk}
where \code{epoce1} and \code{epoce2} are objects inheriting from the \code{epoce} class. 
		\section{Illustrating examples}
		\label{examples}
The package \pkg{frailtypack} provides various functions for models for correlated outcomes and survival data. The Cox proportional hazard model, the shared frailty model for recurrent events (clustered data), the nested frailty model, the additive frailty model and the joint frailty model for recurrent events and a terminal event have already been illustrated elsewhere \citep{Rondeau2005, Rondeau2012frailtypack}.

In this section we focus on extended models for correlated data presented in Section~\ref{joint_models} using three datasets: \code{readmission}, \code{dataMultiv} and \code{colorectal}. Although the joint frailty model has already been presented in the literature, we illustrate its usage as the form of the function has developed in the meantime.

\subsection{Example on dataset readmission for joint frailty models}
\label{readmission}
Dataset \code{readmission} comes from a rehospitalization study of patients after a surgery and diagnosed with colorectal cancer  \citep{Gonzalez2005,Rondeau2012frailtypack}. It contains information on times (in days) of successive hospitalizations and death (or last registered time of follow-up for right-censored patients) counting from date of surgery, and patients characteristics: type of treatment, sex, Dukes' tumoral stage, comorbidity Charlson's index and survival status. The dataset includes 403 patients with 861 rehospitalization events in total. Among the patients 112 (28\%) died during the study.

\subsubsection{Standard joint frailty model}
We adapt the joint model for recurrent events and a terminal event using the gap timescale from the example given in \citep{Rondeau2012frailtypack}. The model \code{modJoint.gap} is defined:
\begin{CodeChunk}
\begin{CodeInput}
R> library("frailtypack")
R> data("readmission")
R> modJoint.gap <- frailtyPenal(Surv(time, event) ~ cluster(id) + dukes +
+    charlson + sex  + chemo + terminal(death), 
+    formula.terminalEvent = ~ dukes + charlson + sex + chemo,
+    data = readmission, n.knots = 8, kappa = c(2.11e+08, 9.53e+11))
\end{CodeInput}
\end{CodeChunk}
This model includes Dukes's stage, Charlson's index, sex and treatment as covariates for both hospitalizations and death. The frailties were from the Gamma distribution (default option) and the baseline hazard functions were approximated by splines with 8 knots and the smoothing parameters for the penalized log-likelihood were $2.11e+8$ for the recurrent part and $9.53e + 11$ for the terminal part. To find the optimal number of knots, we fitted the model a small number of knots (\code{n.knots = 4}) and increased the number of knots until the graph of the baseline hazard functions was not changing importantly anymore. The smoothing parameters were obtained from a shared frailty and Cox models with respectively recurrent and terminal event as the outcome using the cross-validation method.

With this model, it was found that the chemotherapy is a prognostic factor only on death with a positive association (HR$=2.99$, $p<0.001$). Both Charlson's index (Index $\geq3$ vs. Index 0) and Dukes' stage (Stage C and D vs. Stages A, B) were positively related to the recurrent and terminal events. A detailed description of the output for the standard joint frailty models were presented in \citet{Rondeau2012frailtypack}.	

\begin{figure}[h!]
\begin{center}
	\includegraphics[ trim = 0.0cm 0.4cm 0.2cm 0.7cm, clip ]{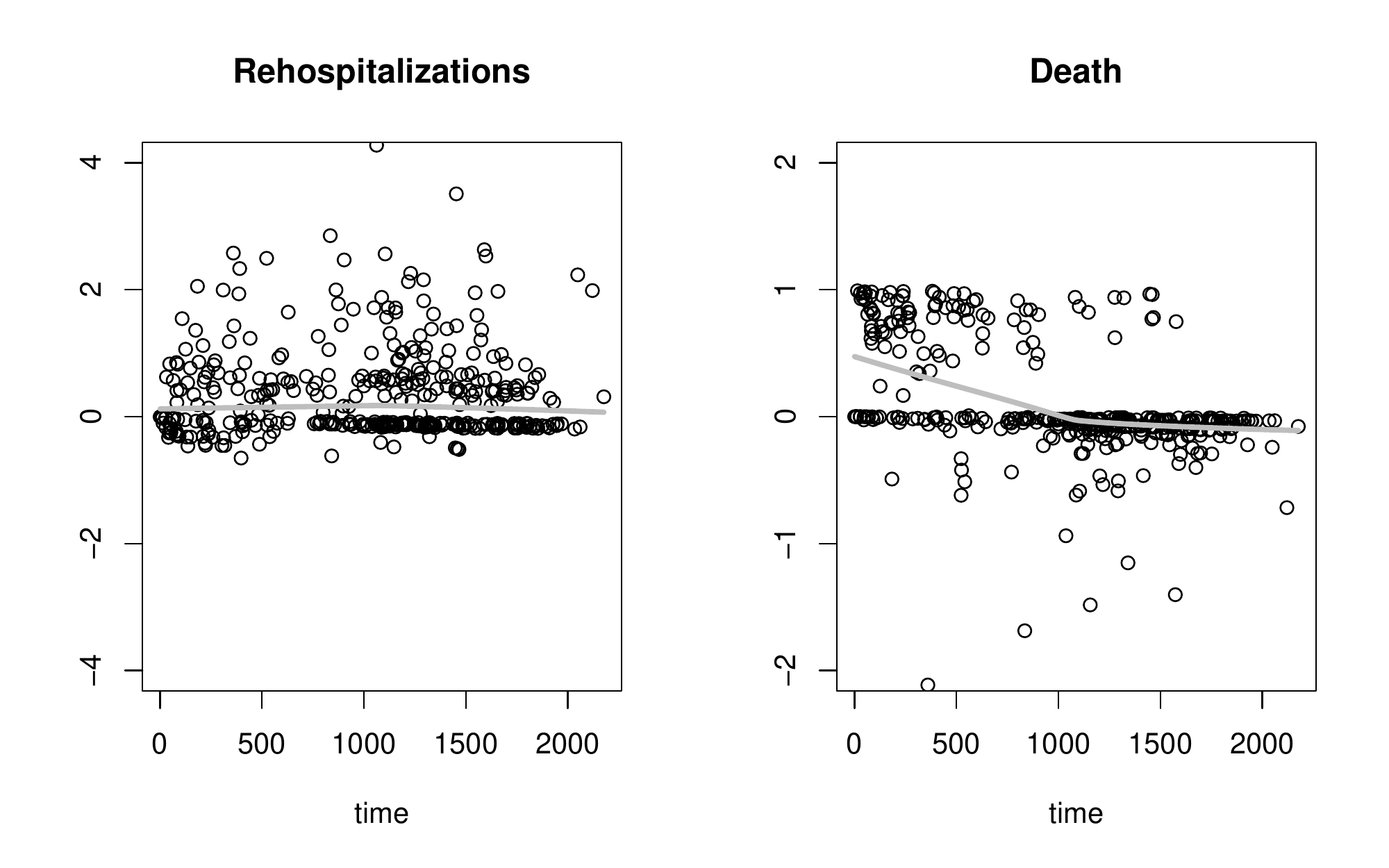} 
	\caption{Martingale residuals for rehospitalizations and death against the follow-up time (in days). The grey line corresponds to a smooth curve obtained with \code{lowess}.}
	\label{figure:residuals1}
	\end{center}
\end{figure}

To verify whether the model predicts correctly the number of observed events, we represent the martingale residuals for both events against the follow-up time. These residuals in a well adjusted model should have a mean equal to 0 and thus a smoothing curve added to a graph should be approximately overlapping with the horizontal line $y=0$. The following  code produces graphs represented in Figure~\ref{figure:residuals1}:
\begin{CodeChunk}
\begin{CodeInput}
R> plot(aggregate(readmission$t.stop, by = list(readmission$id), 
+    FUN = max)[2][ ,1], modJoint.gap$martingale.res, ylab = "",
+    xlab = "time",  main = "Rehospitalizations", ylim = c(-4, 4))
R> lines(lowess(aggregate(readmission$t.stop, by = list(readmission$id),
+    FUN = max)[2][ ,1], modJoint.gap$martingale.res, f = 1), lwd = 3, 
+    col = "grey")
R> plot(aggregate(readmission$t.stop, by = list(readmission$id), 
+    FUN = max)[2][ ,1], modJoint.gap$martingaledeath.res, ylab = "", 
+    xlab = "time", main = "Death", ylim = c(-2, 2))
R> lines(lowess(aggregate(readmission$t.stop, by = list(readmission$id), 
+    FUN = max)[2][ ,1], modJoint.gap$martingaledeath.res, f = 1), lwd = 3, 
+    col = "grey")
\end{CodeInput}
\end{CodeChunk}

For the rehospitalization process the mean of residuals was approximately 0 with the smooth curve close to the line $y=0$, but in case of death this tendency was deviated by relatively higher values for short follow-up times. This may suggest, that the model may have underestimated the number of death in the early follow-up period. The identified individuals of which the residuals result in non-zero mean, had short intervals between their rehospitalization and death (1 day). Indeed, the removal of these individuals (50 patients) resulted in residuals with the mean close to 0 all along the follow-up period (plot not shown here). 

The package \pkg{frailtypack} provides also the estimation of the random effects. Vector \code{frailty.pred} from \code{jointPenal} object contains the individual empirical Bayes estimates. They can be graphically represented for each individual with additional information on number of events (point size) to identify the outlying data. 
\begin{CodeChunk}
\begin{CodeInput}
R> plot(1 : 403, modJoint.gap$frailty.pred, xlab = "Id of patients",
+    ylab = "Frailty predictions for each patient", type = "p",  axes = F,
+    cex = as.vector(table(readmission$id)), pch = 1, ylim = c(-0.1, 7), 
+    xlim = c(-2, 420))
R> axis(1, round(seq(0, 403, length = 10), digit = 0))
R> axis(2, round(seq(0, 7, length = 10), digit = 1))
\end{CodeInput}
\end{CodeChunk}

\begin{figure}[h!]
\begin{center}
	\includegraphics[ trim = 0.0cm 0.6cm 0.2cm 1.4cm, clip ]{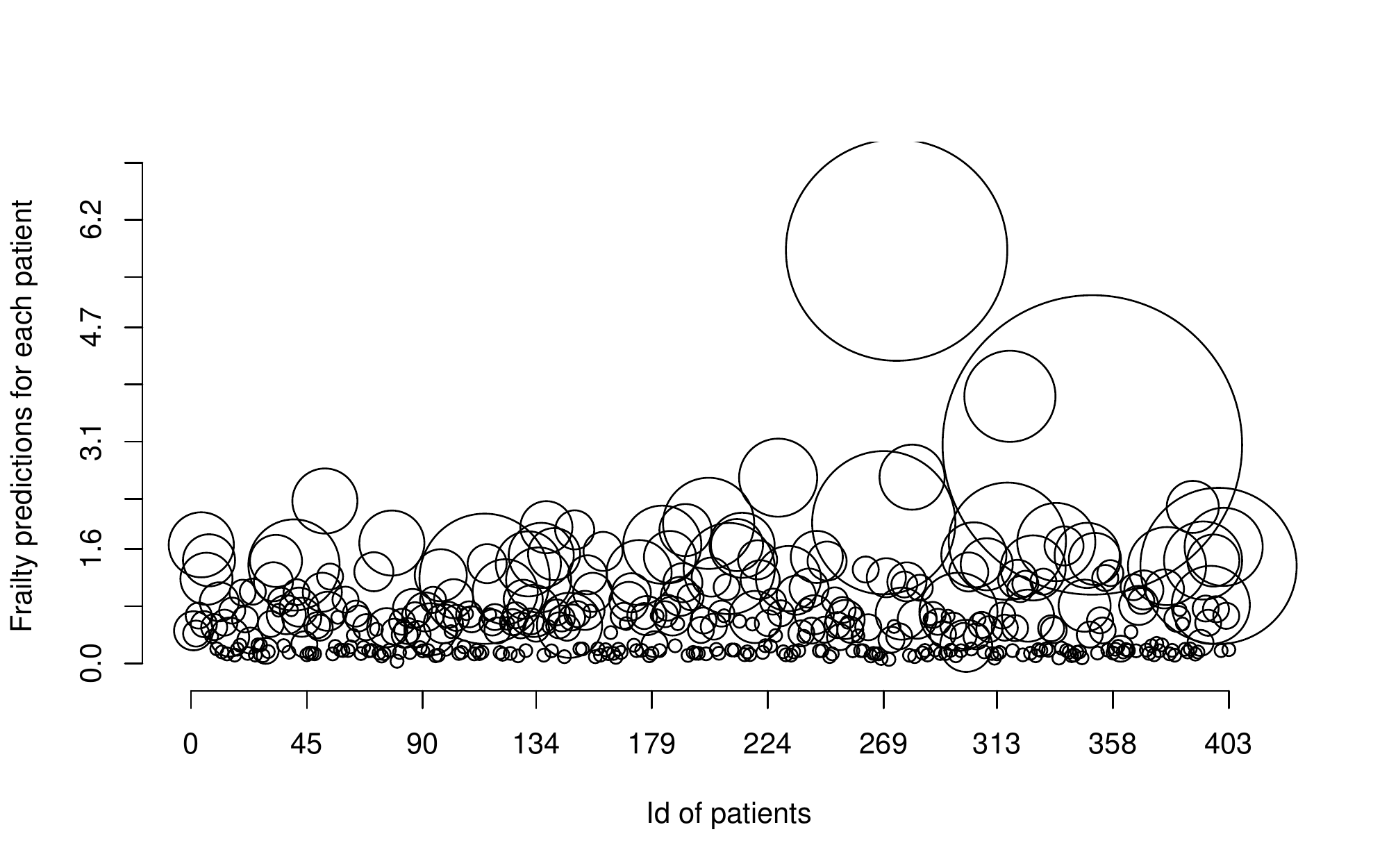} 
	\caption{Individual prediction of the frailties. The size of points corresponds to  number of an individual's recurrent events.}
	\label{figure:frailty1}
	\end{center}
\end{figure}

Figure~\ref{figure:frailty1} shows the values of frailty prediction for each patient with an association to the number of events (the bigger the point, the greater the number of rehospitalizations). The frailties tended to have bigger values if the number of events of a given individual was high. From the plot it can be noticed that there was an outlying frailty suggesting verification of the follow-up of the concerned individual. 

\subsubsection{Time-varying coefficients}

In the framework of standard joint frailty models, it is possible to fit the models with time-varying effects of prognostic factors. Using function \code{timedep} in a formula of \code{frailtyPenal}, the time-dependent coefficients can be estimated using B-splines of order $q$ (option \code{betaorder}) with $m$ interior knots (option \code{betaknots}). In the example of \code{readmission} dataset we are interested in verifying whether the variable sex has a time-varying effect on both recurrent and terminal events. Thus, we fit a model equivalent to \code{modJoint.gap} but with time dependent effects assuming quadratic B-splines and 3 interior knots:
\begin{CodeChunk}
\begin{CodeInput}
R> modJoint.gap.timedep <- frailtyPenal(Surv(time, event) ~ cluster(id) + 
+    dukes + charlson + timedep(sex) + chemo + terminal(death), 
+    formula.terminalEvent = ~ dukes + charlson + timedep(sex) + chemo,
+    data = readmission, n.knots = 8, kappa = c(2.11e+08, 9.53e+11),
+    betaorder = 3, betaknots = 3)
\end{CodeInput}
\end{CodeChunk}

In the result, using the method \code{print} the estimated values of parameters with time-constant effects and graphics of log-hazard ratios for time-dependent variables for each events are obtained (Appendix B, Figure~\ref{figure:timedep}). For rehospitalizations, we found firstly a protective effect for females ($\beta(t)<0$) and later an increased risk ($\beta(t)>0$). For death, at the beginning, the effect of sex was weakening the risk but shortly became non-influential ($\beta(t)$ around 0). 

The PH assumption for the variable sex can be checked using the LR test. We compare two models: \code{modJoint.gap} (related to the null hypothesis that the effect is constant in time: $H_0\ :\ \beta_{R}(t)=\beta_R,\ \beta_T(t)=\beta_T$) and \code{modJoint.gap.timedep} (related to the alternative hypothesis of time-varying effects is time-varying: $H_1\ :\ \beta_R(t)\ne \beta_R,\ \beta_T(t)\ne\beta_T$):
\begin{CodeChunk}
\begin{CodeInput}
R> LR.statistic <- -2 * modJoint.gap$logLik + 2 * modJoint.gap.timedep$logLik
R> p.value <- signif(1 - pchisq(LR.statistic, df = 10), 5) 
\end{CodeInput}
\end{CodeChunk}
Given the obtained p-value=0.049, the PH assumption for the variable sex is not satisfied (at the level 0.05). Next, we checked whether sex with time-dependent effects is an influential prognostic factor. Thus, again, we used the LR test to compare two models: \code{modelJoint.gap.nosex} without the covariate sex (model related to the null hypothesis: $H_0\ :\ \beta_R(t)=0,\ \beta_T(t)=0$) and \code{modelJoint.gap.timedep} (related to the alternative hypothesis: $H_1\ :\ \beta_R(t)\ne 0,\ \beta_T(t)\ne 0$):
\begin{CodeChunk}
\begin{CodeInput}
R> modJoint.gap.nosex <- frailtyPenal(Surv(time, event) ~ cluster(id) + 
+    dukes + charlson + chemo + terminal(death), formula.terminalEvent = ~ 
+    dukes + charlson + chemo, data = readmission, n.knots = 8, 
+    kappa = c(2.11e+08, 9.53e+11))
R> LR.statistic <- -2 * modJoint.gap.nosex$logLik + 
+    2 * modJoint.gap.timedep$logLik
R> p.value <- signif(1 - pchisq(LR.statistic, df = 12), 5)
\end{CodeInput}
\end{CodeChunk}
The test showed that the sex variable had a significant time-varying effect (p-value$<0.001$). In this example, the variable sex was found significant for both PH model and non-PH model, but we showed that this variable did not satisfy the PH assumption.

\subsubsection{Dynamic predictions}
Using the joint model \code{modJoint.gap} we can calculate  predicted probabilities of death using \code{prediction}. These predictions may serve as a tool to compare two exemplars of patients with different history of recurrences but the same values of the prognostic factors to study the effect of the events on the survival \citep{Mauguen2013}. On the other hand, the influence of some explanatory covariates can be examined for patients that are considered to have the same histories of recurrences. Here, we aimed at evaluating the predictive effect of the Dukes' stage on survival considering the history of hospitalizations. We compared the predicted risk of death for two patients having the same characteristics (men with Charlson's index 0 and the chemotherapy treatment) and having two hospitalizations: 1 and 1.5 year after their surgeries. We set the time of prediction for 2 years and calculate the probability in a time window  of 3 years (we apply a moving window with a step of 0.5 year). Patient 1 had Dukes' stage A and patient 2, Dukes' stage D. We focused on the predicted probabilities regarding the complete history of recurrences (Equation~\ref{prediction}) and compared the results with those obtained using the incomplete history and ignoring the history.

To prepare data for the predictions, we started with an empty data frame with the variables of interest and the covariates:
\begin{CodeChunk}
\begin{CodeInput}
R> datapred <- data.frame(time = 0, event = 0, id = 0, dukes = 0, charlson = 
+  0, sex = 0, chemo = 0)
R> datapred[ ,4 : 7] <- lapply(datapred[ , 4 : 7],as.factor)
R> levels(datapred$dukes) <- c(1, 2, 3)
R> levels(datapred$charlson) <- c(1, 2, 3)
R> levels(datapred$sex) <- c(1, 2)
R> levels(datapred$chemo) <- c(1, 2)
\end{CodeInput}
\end{CodeChunk}
Patient 1 with Dukes' stage A had two observed hospitalizations at 365th and 548th day after the surgery:
\begin{CodeChunk}
\begin{CodeInput}
R> datapred[1, ] <- c(365, 1, 1, 1, 1, 1, 2) 
R> datapred[2, ] <- c(548, 1, 1, 1, 1, 1, 2)
\end{CodeInput}
\end{CodeChunk}
Patient 2 had the hospitalizations observed in the same times as Patient 1 but was assumed to have Dukes' stage B:
\begin{CodeChunk}
\begin{CodeInput}
R> datapred[3, ] <- c(365, 1, 2, 3, 1, 1, 2)
R> datapred[4, ] <- c(548, 1, 2, 3, 1, 1, 2)
\end{CodeInput}
\end{CodeChunk}
We calculated the predictions for both patients: 
\begin{CodeChunk}
\begin{CodeInput}
R> pred <- prediction(modJoint.gap, datapred, t = 730, window = 
+    seq(1, 1096, 183), MC.sample = 500)
R> plot(pred, conf.bands = TRUE)
\end{CodeInput}
\end{CodeChunk}
\begin{figure}[h!]
\begin{center}
	\includegraphics[ trim = 0.0cm 0.7cm 0.2cm 0.7cm, clip ]{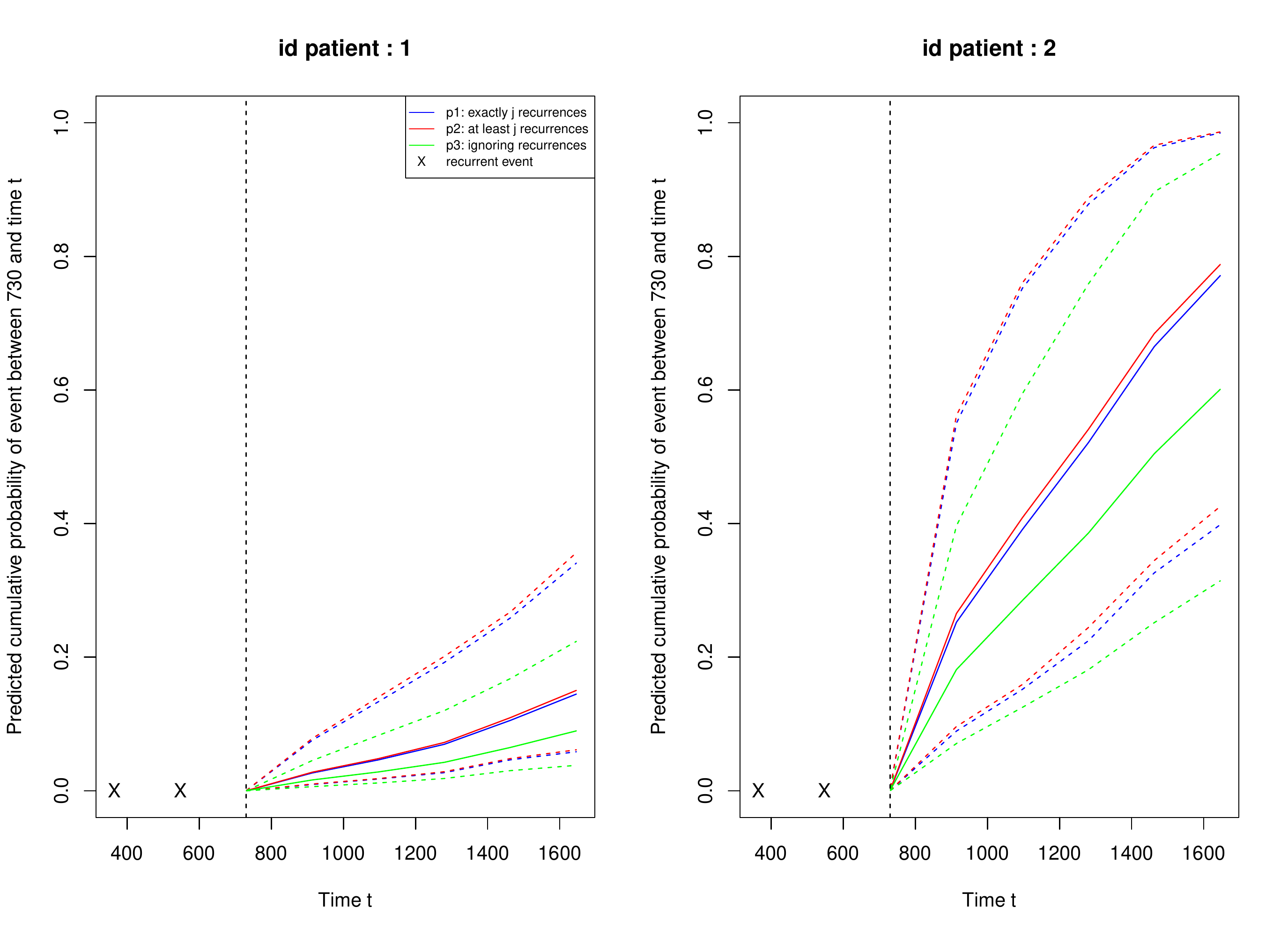} 
	\caption{Predicted probabilities of death for two patients sharing the same history of recurrences and characteristics except of the Dukes' stage. Patient 1 had Dukes' stage A and patient 2, Dukes' stage D. Dashed lines represent the MC confidence intervals.}
	\label{figure:prediction1}
	\end{center}
\end{figure}

In the result, three types of predictions were calculated for 6 time horizons. As it has been already observed from the estimates of the model, an increased Dukes's stage (D) was positively associated with death. Figure~\ref{figure:prediction1} compares the predicted probabilities of death in the three settings. The predictions using the exact number of recurrences (2 hospitalizations) and at least 2 recurrences were very close to each other and were higher than  the risk obtained without the information on recurrent events. Indeed, the significant estimate of variance of the frailty $\theta$ and significant positive estimate of $\alpha$  implied the positive association between the process of recurrences and death. Finally, all the predicted probabilities were higher for the patient with Dukes' stage D compared to the patient with Dukes' stage A.

\subsubsection{Joint frailty model for clustered data}
The joint models for clustered survival data can be estimated using again \code{frailtyPenal} function. A dataset should include information on two survival outcomes for individuals from several groups. This model is presented  using \code{readmission} dataset with artificially created clusters on individuals. The first survival event will be the first observed rehospitalization and the second event, death. The framework of semi-competing risks is used here, thus individuals' follow-up stops at time of the rehospitalization, death or in case when none of these events are observed, the censoring time. We consider 6 clusters defined by a new variable \code{group}:
\begin{CodeChunk}
\begin{CodeInput}
R> readmission <- transform(readmission, group = id %% 6 + 1 )
R> readm_cluster <- subset(readmission, (t.start == 0 & event == 1) 
+    | event == 0) 
\end{CodeInput}
\end{CodeChunk}
New dataset \code{readm_cluster} includes clusters with 97 to 107 individuals per group.
For definition of a clustered joint model two inner functions are required in \code{frailtyPenal}, \code{num.id} for the individual level and \code{cluster} for the groups:
\begin{CodeChunk}
\begin{CodeInput}
R> joi.clus <- frailtyPenal(Surv(t.start, t.stop, event) ~ cluster(group) +
+    num.id(id) + dukes + sex + chemo + terminal(death), 
+    formula.terminalEvent = ~ dukes + sex + chemo, data = readm_cluster,
+    n.knots = 8, kappa = c(1.e+10, 1.e+10), recurrentAG = T, Alpha = "None")
\end{CodeInput}
\end{CodeChunk}
In the result, the estimates of prognostic factors for both types of events are obtained. The estimate of the variance of the frailty term indicates whether, at the cluster level, the processes are associated with each other and measures the heterogeneity between individuals (intra-cluster correlation). In the given example the estimate of the variance $\theta$ was significantly different from 0 (p-value=0.037), thus there was a positive association between the risk of hospitalizations and death via the non-observed frailty.

In case of the joint frailty models for clustered data it should be noted that sufficient amount of information must be provided, ie. number of observation per cluster. Otherwise, given the complexity of the model, the convergence might not be attained. The parameter $\alpha$ is assumed to be equal to 1 as these models are defined in the framework of semi-competing risks and not of recurrent events.

\subsubsection{General joint frailty model}
To estimate the general frailty model, argument \code{jointGeneral} must be equal to \code{TRUE} in function \code{frailtyPenal}. We applied this model to original  \code{readmission} dataset assuming two independent frailty terms using a following code:
\begin{CodeChunk}
\begin{CodeInput}
R> modJoint.general <- frailtyPenal(Surv(time, event) ~ cluster(id) + dukes +
+    charlson + sex + chemo + terminal(death), formula.terminalEvent = ~ 
+    dukes + charlson + sex + chemo, data = readmission, jointGeneral = TRUE, 
+    n.knots = 8, kappa = c(2.11e+08, 9.53e+11))
\end{CodeInput}
\end{CodeChunk}

In the output of the function, estimations of the variances of both random effects are given. For the analyzed example, the estimated variance $\theta$  of the frailty $u_i$ associating recurrent events and death indicates strong relationship between the processes ($\hat{\theta}=0.68$, p-value < 0.001). Moreover, the estimate of $\eta$  implies small but significant dependence between the recurrent event gap times explained by the frailty $v_i$ ($\hat{\eta}=0.01$, p-value < 0.001). This information complements the inference from the standard joint frailty model because it separates the correlation linked to the recurrent events with correlation between the recurrent events and the terminal event.

\subsection{Example on dataset dataMultiv for multivariate joint frailty model}
For the following example we applied a generated dataset for 800 individuals from Model (\ref{joint2recsurv}) with 2 types of recurrent events and a terminal event. The random effects were assumed to follow the normal distribution $u_i\sim\mathcal{N}(0,0.5)$, $v_i\sim\mathcal{N}(0,0.5)$ and the correlation coefficient $\rho=0.5$. The coefficients for the random effects were $\alpha_1=\alpha_2=1$. The baseline hazard functions $r_0^{(1)}(t)$, $r_0^{(2)}(t)$ and $\lambda_0(t)$ followed the Weibull distribution and the time for right censoring was fixed at 5. The generated data included 1652 observations. For detailed description of the generation scenario see \citet{Mazroui2012}.

The dataset includes individuals times of events with variables indicating the type of event: \code{INDICREC} for the recurrent event of type 1 (local recurrences), \code{INDICMETA} for the recurrent event of type 2 (metastases) and \code{INDICDEATH} for censoring status (death). Additionally there are 3 binary covariates \code{v1}, \code{v2} and \code{v3}.

\subsubsection{Multivariate frailty model}

We consider the multivariate frailty model for the exemplary dataset \code{dataMultiv} to study jointly local recurrences, metastases and death for patients diagnosed with cancer. To define the model, three formulas must be defined in the function with additional indication on variables including status of the second recurrent event (\code{event2}) and of the terminal event (\code{terminal}), both included in the first formula. All the baseline hazard functions must be of the same type (Weibull, splines or piecewise constant). We fit the model as follows  (computational time 54 minutes on a personal computer with an Intel Core i7 3.40 GHz processor and 8 GB RAM runing Windows 7):
\begin{CodeChunk}
\begin{CodeInput}
R> data("dataMultiv")
R> modMultiv.spli <- multivPenal(Surv(TIMEGAP, INDICREC) ~ cluster(PATIENT) +  
+    v1 + v2 + event2(INDICMETA) + terminal(INDICDEATH), formula.Event2 = ~ 
+    v1 + v2 + v3, formula.terminalEvent = ~ v1, data = dataMultiv, 
+    n.knots = c(8, 8, 8), kappa = c(1, 1, 1), initialize = FALSE)
\end{CodeInput}
\end{CodeChunk}
Option \code{initialize} indicates whether initialize the parameters (including parameters for the baseline hazard functions) using the estimates of separate models: shared frailty models (for the two types of recurrent events) and a Cox proportional hazard model (for the terminal event). The output of function \code{multivPenal} is presented below:
\begin{CodeChunk}
\begin{CodeInput}
R> modMultiv.spli
\end{CodeInput}
\end{CodeChunk}
\begin{CodeChunk}
\begin{CodeOutput}
Call:
multivPenal(formula = Surv(TIMEGAP, INDICREC) ~ cluster(PATIENT) + 
    v1 + v2 + event2(INDICMETA) + terminal(INDICDEATH), formula.Event2 = ~
    v1 + v2 + v3, formula.terminalEvent = ~v1, data = dataMultiv, 
    initialize = FALSE, n.knots = c(8, 8, 8), kappa = c(1, 1, 1))
    
  Multivariate joint gaussian frailty model for two survival outcomes
  and a terminal event 
  using a Penalized Likelihood on the hazard function 
  
Recurrences 1:
------------ 
       coef exp(coef) SE coef (H) SE coef (HIH)       z          p
v1 0.565676   1.76064    0.111603      0.111638 5.06863 4.0068e-07
v2 0.631891   1.88117    0.106534      0.106519 5.93138 3.0040e-09

Recurrences 2:
------------- 
        coef exp(coef) SE coef (H) SE coef (HIH)        z          p
v1  0.837140  2.309752    0.127631      0.127554  6.55905 5.4152e-11
v2 -0.641487  0.526509    0.127111      0.127075 -5.04668 4.4956e-07
v3  0.312774  1.367212    0.118103      0.118057  2.64832 8.0892e-03

Terminal event:
---------------- 
       coef exp(coef) SE coef (H) SE coef (HIH)       z          p
v1 0.367778   1.44452   0.0987691     0.0984928 3.72362 0.00019639

 Parameters associated with Frailties: 
   theta1 : 0.523131 (SE (H): 0.537725 ) p = 0.16531 
   theta2 : 0.25968 (SE (H): 0.966704 ) p = 0.39411 
   alpha1 : 0.54705 (SE (H): 0.111603 ) p = 9.4993e-07 
   alpha2 : 0.595186 (SE (H): 0.106534 ) p = 2.3125e-08 
   rho : 0.738084 (SE (H): 0.0987691 ) 
 
   penalized marginal log-likelihood = -594.7
   LCV = the approximate likelihood cross-validation criterion
         in the semi parametric case     = 0.477466 
         
   n= 1318
   n recurrent events of type 1= 518  n subjects= 800
   n recurrent events of type 2= 334
   n terminal events= 636
   number of iterations:  16 
   
   Exact number of knots used:  8   8   8 
   Value of the smoothing parameters:  1 1 1
\end{CodeOutput}
\end{CodeChunk}
The output presents the results for prognostic factors estimates for each type of event. The estimates of parameters associated with the random effects are given by the variance of the frailty related to the first type of the recurrent events and the association with the terminal event (\code{theta1}), the variance of the frailty related to the second type of the recurrent events and the association with the terminal event (\code{theta2}) and the correlation coefficient between the frailties (\code{rho}). The sign and strength of the dependency between the recurrent event of type 1 (2) and the terminal event is represented by \code{alpha1} (\code{alpha2}). In the analyzed example, both \code{theta1} and \code{theta2} were not significantly different from 0, thus there were no dependencies between the processes explained by the non-observed factors.

\subsection{Example on dataset colorectal for models with longitudinal data}
\label{colorectal}
Datasets \code{colorectal} and \code{colorectal.Longi} represent a random selection of 150 patients from a multi-center randomized phase III clinical trial FFCD 2000-05 of patients diagnosed with metastatic colorectal cancer not amenable to curative intent surgery.  The trial was conducted between 2002 and 2007 in France by F\'{e}d\`{e}ration Francophone de Canc\'{e}rologie Digestive (FFCD) \citep{Ducreux2011}. The data contains a follow-up 
 of tumor size measure (sum of the longest diameters of target lesions) and times of apparition of new lesions as recurrent events. Moreover, some baseline characteristics (age, WHO performance status and previous resection), treatment arm (combination vs. sequential)  and time of death (or last observed time for a right-censored individual) are included in the data. Dataset \code{colorectal} provides information on recurrent event and death and dataset \code{colorectal.Longi} on the measurements of tumor size. A total of 906 tumor size measurements and 289 of recurrences were recorded for patients included. Among them, 121 died during the study.

The variable \code{tumor.size} in \code{colorectalLongi} is the transformed sum of the longest diameters ($SLD^*$) of an individual's target lesions measured during a visit ($SLD^*=(SLD^{0.3}-1)/0.3$). The status of new lesions occurrence is registered in \code{new.lesions} in dataset \code{colorectal}. In this dataset start of time interval \code{time0} (0 or time of previous recurrence) and time of event \code{time1} (recurrence or censoring by terminal event) represent information for times of apparition of new lesions and for death (or right censoring). 

We provide extracts of \code{colorectal} and \code{colorectalLongi} datasets to guide the users how to prepare suitable dataset for joint models with longitudinal data using the package. These functions require datasets in long format (one row per observation and usually several rows per individuals) and only in case of \code{data} in \code{longiPenal} the long format is one row per individual (as it contains information on termianl event only). In example of the colorectal dataset this is represented as follows:
\begin{Sinput}
R> data("colorectalLongi")
R> head(colorectalLongi, 10)
\end{Sinput}
\begin{Soutput}
   id      year tumor.size treatment         age who.PS prev.resection
1   1 0.0000000  5.2276794         S 60-69 years      0             No
2   1 0.2131147  4.4926205         S 60-69 years      0             No
3   1 0.4590164  4.6000876         S 60-69 years      0             No
4   1 0.6311475  4.5333227         S 60-69 years      0             No
5   2 0.0000000  3.0454011         C   >69 years      0             No
6   2 0.1639344  1.3919052         C   >69 years      0             No
7   2 0.2814208  1.2063562         C   >69 years      0             No
8   2 0.4316940  1.2063562         C   >69 years      0             No
9   2 0.5846995  0.9462067         C   >69 years      0             No
10  2 0.7377049  1.9353985         C   >69 years      0             No
\end{Soutput}
\begin{Sinput}
R> data("colorectal")
R> head(colorectal, 5)
\end{Sinput}
\begin{Soutput}
  id     time0     time1 new.lesions treatment         age who.PS
1  1 0.0000000 0.7095890           0         S 60-69 years      0
2  2 0.0000000 1.2821918           0         C   >69 years      0
3  3 0.0000000 0.5245902           1         S 60-69 years      1
4  3 0.5245902 0.9207650           1         S 60-69 years      1
5  3 0.9207650 0.9424658           0         S 60-69 years      1
  prev.resection state   gap.time
1             No     1 0.70958904
2             No     1 1.28219178
3             No     0 0.52459017
4             No     0 0.39617486
5             No     1 0.02170073
\end{Soutput}
\subsubsection{Joint model for longitudinal data and a terminal event}

Firstly, we estimated the bivariate joint model for longitudinal data and a terminal event (\ref{srem}). We considered a left-censored biomarker, transformed tumor size measurements are not observed below a threshold -3.33 (which corresponds to 'zero' measures in the nontransformed data). The value of  $\kappa$ smoothing parameter was chosen using a corresponding reduced model, ie. a Cox model for the terminal event. For a model with a random intercept and slope for the biomarker and the link function being the random effects of the biomarker we used the following form of the \code{longiPenal} function using \code{colorectalLongi} dataset and a subset \code{colorectal} with only information on the terminal event (\code{colorectalSurv}):
\begin{CodeChunk}
\begin{CodeInput}
R> colorectalSurv <- subset(colorectal, new.lesions == 0)
R> modLongi <- longiPenal(Surv(time1, state) ~ age + treatment + who.PS
+   + prev.resection, tumor.size ~  year * treatment  + age + who.PS,
+   colorectalSurv, data.Longi = colorectalLongi, random = c("1", "year"),
+   id = "id", link = "Random-effects", left.censoring = -3.33,
+   n.knots = 8, kappa = 0.93, method.GH = "Pseudo-adaptive", n.nodes = 7)
R> modLongi 
\end{CodeInput}
\end{CodeChunk}
\begin{CodeChunk}
\begin{CodeOutput}
> modLongi
Call:
longiPenal(formula = Surv(time1, state) ~ age + treatment + who.PS + 
    prev.resection, formula.LongitudinalData = tumor.size ~ year * treatment
    + age + who.PS, data = colorectalSurv, data.Longi = colorectalLongi, 
    random = c("1", "year"), id = "id", link = "Random-effects", 
    left.censoring = -3.33, n.knots = 8, kappa = 0.93, method.GH = 
    "Pseudo-adaptive", n.nodes = 7)

  Joint Model for Longitudinal Data and a Terminal Event 
  Parameter estimates using a Penalized Likelihood on the hazard function 
  and assuming left-censored longitudinal outcome 
  Association function: random effects 

Longitudinal outcome:
---------------- 
                     coef SE coef (H) SE coef (HIH)         z          p
Intercept        3.023259    0.187384      0.187384 16.134028     <1e-16
year            -0.303105    0.135847      0.135847 -2.231218 2.5667e-02
treatmentC       0.048720    0.213961      0.213961  0.227703 8.1988e-01
age60-69 years  -0.017015    0.168804      0.168804 -0.100797 9.1971e-01
age>69 years    -0.294597    0.131867      0.131867 -2.234047 2.5480e-02
who.PS1          0.106983    0.116952      0.116952  0.914761 3.6032e-01
who.PS2          0.739941    0.175652      0.175652  4.212540 2.5251e-05
year:treatmentC -0.634663    0.183640      0.183640 -3.456020 5.4822e-04

          chisq df global p
age     5.00113  2 8.20e-02
who.PS 18.58228  2 9.22e-05

Terminal event:
------------- 
                       coef exp(coef) SE coef (H) SE coef (HIH)         z 
age60-69 years    -0.226281  0.797494    0.243004      0.243004 -0.931181 
age>69 years      -0.100922  0.904004    0.223801      0.223801 -0.450944 
treatmentC        -0.090385  0.913580    0.198843      0.198843 -0.454554 
who.PS1           -0.116794  0.889769    0.218352      0.218352 -0.534889 
who.PS2            0.802399  2.230887    0.258302      0.258302  3.106433 
prev.resectionYes -0.225774  0.797898    0.193418      0.193418 -1.167288 

         p
3.5176e-01
6.5203e-01
6.4943e-01
5.9273e-01
1.8936e-03
2.4309e-01

         chisq df global p
age    1.07045  2  0.58600
who.PS 9.93603  2  0.00696

Components of Random-effects covariance matrix B1: 
                             
Intercept  1.972030 -0.519742
year      -0.519742  0.943568

Association parameters: 
               coef        SE       z         p
Intercept 0.3203431 0.0848202 3.77673 0.0001589
year      0.0432288 0.1668875 0.25903 0.7956100

Residual standard error:  0.954237  (SE (H):  0.027079 ) 
 
      penalized marginal log-likelihood = -1684.01
      Convergence criteria: 
      parameters = 0.000127 likelihood = 0.000361 gradient = 1.47e-08 

      LCV = the approximate likelihood cross-validation criterion
            in the semi parametrical case     = 1.62312 

      n= 150
      n repeated measurements= 906
      Percentage of left-censored measurements= 3.75 %
      Censoring threshold s= -3.33
      n events= 121
      number of iterations:  26 

      Exact number of knots used:  8 
      Value of the smoothing parameter:  0.93
      Gaussian quadrature method:  Pseudo-adaptive with 7 nodes 
\end{CodeOutput}
\end{CodeChunk}
On average, the tumor size significantly decreased in time in interaction with the treatment, this effect was more important in the C arm (-0.63, p-value = 0.001). However, there was no effect of treatment arm on the risk of death (HR = 0.91, p-value = 0.65). The age of patients at baseline did not have any effect neither on the tumor size nor on the risk of death. The performance status WHO 2 evaluated before the treatment was a prognostic factor both for the tumor size (0.74, p-value<0.001) and overall survival (HR = 2.23, p-value = 0.001). It should be noted that the model was fitted on the subset of the original trial in which, using the data of all patients, some of the prognostic effects were found different \citep{Krol2015}.

The processes were linked together via the random intercept and slope of the longitudinal trajectory. This association was significant for the random intercept  implying that with the increase of individual deviation from the population average tumor size, the risk of death increased as well ($\hat{\eta}_{t1} = 0.32$, p-value = 0.002).

To verify the goodness-of-fit of the model, the estimated baseline hazard function, martingale residuals for the terminal event and residuals for the longitudinal outcome can be plotted using following code (results not shown here):
\begin{CodeChunk}
\begin{CodeInput}
R> plot(modLongi, main = "Hazard function")
R> plot(aggregate(colorectalSurv$time1, by = list(colorectalSurv$id), 
+    FUN = max)[2][ ,1], modLongi$martingaledeath.res, ylab = "", xlab = 
+    "time", main = "Martingale Residuals - Death", ylim = c(-4.2, 4.2))
R> lines(lowess(aggregate(colorectalSurv$time1, by = list(colorectalSurv$id), 
+    FUN = max)[2][ ,1], modLongi$martingaledeath.res, f = 1), lwd = 3, 
+    col = "grey")
R> qqnorm(modLongi$marginal_chol.res, main = "Marginal Cholesky residuals", 
+    xlab = "")
R> qqline(modLongi$marginal_chol.res)
R> plot(modLongi$pred.y.cond, modLongi$conditional.res, xlab = "Fitted",
+    ylab = "Conditional residuals", main = "Conditional Residuals 
+    vs. Fitted Values")
\end{CodeInput}
\end{CodeChunk}

Next, we explored predictiveness of this model comparing it with a model in which the link function was represented by the current level of the biomarker $m_i(t)$. Thus, we fitted a bivariate model with the same covariates and characteristics but with the option \code{link =  "Current-level"} (model \code{modLongi2}). Then, for times of prediction between 1.2 and 2.5 years we calculated the EPOCE estimator. We compared the models by plotting values of CVPOL$_a$ and tracking intervals for the differences between the models. 
\begin{CodeChunk}
\begin{CodeInput}
R> modLongi2 <- longiPenal(Surv(time0, time1, state) ~ age + treatment +
+    who.PS + prev.resection, tumor.size ~  year * treatment + age + who.PS, 
+    colorectalSurv, data.Longi = colorectalLongi, random = c("1", "year"), 
+    id = "id", link = "Current-level", left.censoring = -3.33, 
+    n.knots = 8, kappa = 0.93, method.GH = "Pseudo-adaptive", n.nodes = 7)
R> time <- seq(1.2, 2.4, 0.1)
R> epoce <- epoce(modLongi, time)
R> epoce2 <- epoce(modLongi2, time)
R> diff <- Diffepoce(epoce, epoce2)
R> plot(temps, epoce$cvpol, ylab = "CVPOL", xlab = "time", pch = 6, 
+    col = "darkcyan", type= "b", ylim = c(0.4, 1.0))
R> points(temps, epoce2$cvpol, pch = 15, col = 'brown3',  type = "b", lty = 2)
R> legend(1.9, 1, legend = c("modLongi", "modLongi2"), pch = c(6, 15), 
+    col = c("darkcyan", "brown3"), bty = "n", cex = 0.7, lty = c(1, 2))
R> plot(diff)
\end{CodeInput}
\end{CodeChunk}
\begin{figure}[h!]
\begin{center}
	\includegraphics[ trim = 0cm 0.5cm 1cm 1.5cm, clip, width=11cm]{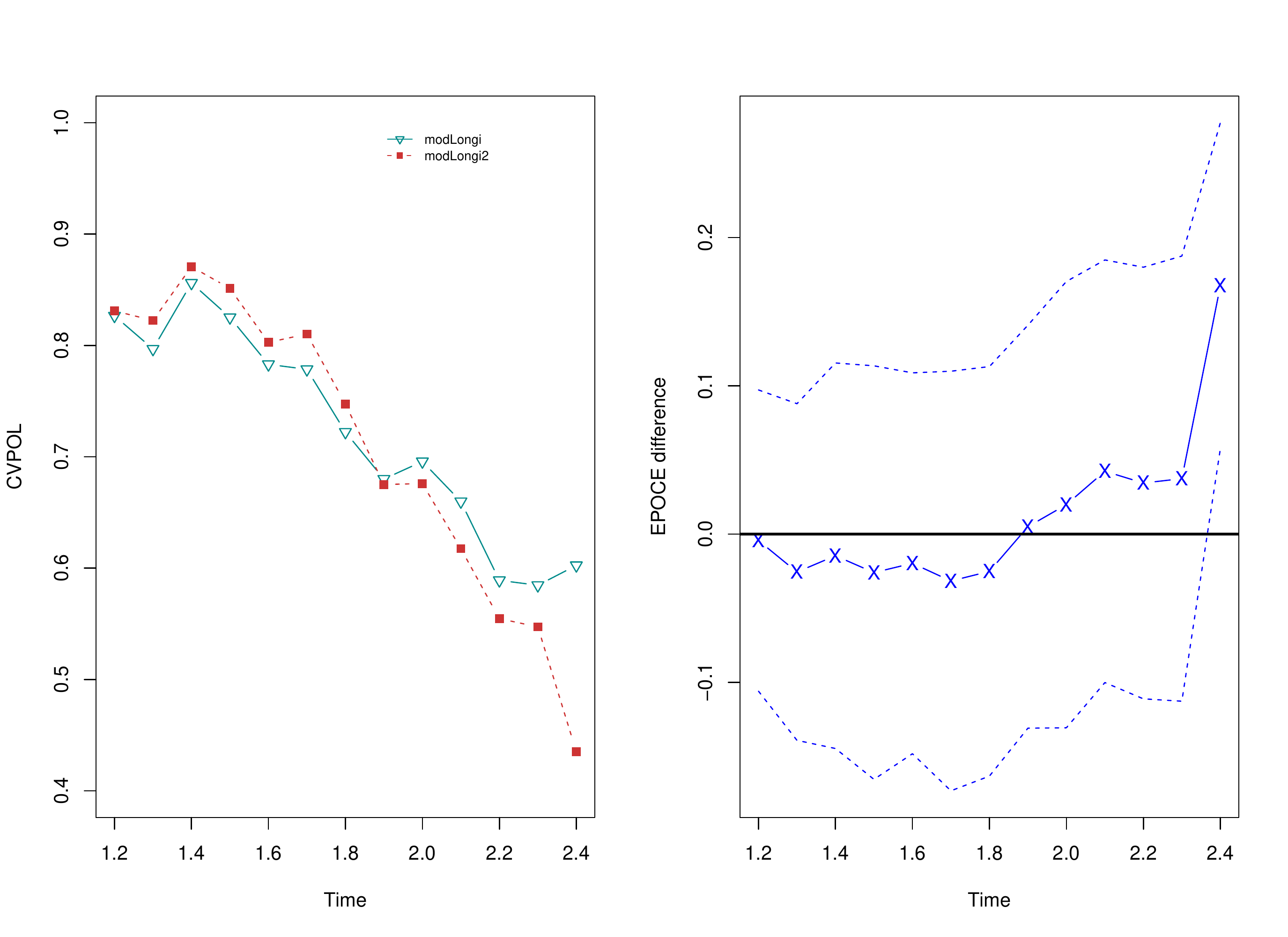} 
	\caption{Estimated cross-validated EPOCE computed from the joint models until 2.5 years (left picture) and differences in the estimates with 95\% tracking intervals (right picture).}
	\label{figure:epoce2}
	\end{center}
\end{figure}

The results of CVPOL$_a$ are presented in Figure~\ref{figure:epoce2}. Model \code{modLongi} with the random effects as the link function  had better predictive abilities than model \code{modLongi2} with  the current level of the biomarker as the link function until 1.9 year. After this time point, the tendency inversed. However, these difference were significant only at 2.4 years of treatment. Using method \code{AtRisk} from \code{epoce} object, we may verify for how many subjects at risk, CVPOL$_a$ was calculated at this prediction time (\code{epoce$AtRisk}). The significant difference between the models was found for the time point for which the number of subjects considered was relatively small (25). Thus, we can conclude that the models are close to each other and the choice of the link function does not strongly influence the predictiveness.

The comparison of the predictive abilities of bivariate models can be useful for the choice of the appropriate trivariate model. If we are interested in the prediction of death  it is important to choose a model that provide the best predictive abilities for the survival. For this reason, we choose a model with such a structure of the link function that provide the best predictive abilities, for both bivariate and trivariate model. First, we fit appropriate bivariate models (without the recurrent part) with different link structures and compare their predictive abilities. Then, we fit the trivariate model with the same link structure as in the bivariate model with the best predictive abilities. For the colorectal dataset the differences in CVPOLa were not of great importance, thus we apply a trivariate model with the random effects of the biomarker as the link function, as this model is less computationally intensive.

\subsubsection{Joint model for longitudinal data, recurrent events and a terminal event}

The package \pkg{frailtypack} allows the estimation of the trivariate models for a longitudinal biomarker, recurrent events and a terminal event. In the dataset colorectal, the recurrent events are represented by the appearance of new lesions during the treatment. Usually in clinical trials the size of new lesions is not registered and thus, their burden cannot be simply added to the measure of the tumor size of the target lesions. However, it is of interest to add the information on new lesions to a model as it influences overall survival. A trivariate model can be a solution for such a goal and can be implemented using function \code{trivPenal}.

We fitted a model with the calendar timescale for recurrent events and the baseline hazard functions approximated by splines with 8 knots. The smoothing parameter values were found from the separate models (shared frailty model and Cox model). A random intercept and a random slope and the left-censoring (threshold $s=-3.33$) were assumed for the biomarker. The pseudo-adaptive Gaussian quadrature with 9 nodes was chosen for calculation of integrals. 

Firstly, we found initial values for the covariates. For the longitudinal outcome and the terminal event we used the estimates from the bivariate model, \code{modLongi}. The covariates related to the recurrent events initialized using the results of a shared frailty model for the appearance of new lesions:
\begin{CodeChunk}
\begin{CodeInput}
modShared <- frailtyPenal(Surv(time0, time1, new.lesions) ~ cluster(id) + 
+    age + treatment + who.PS, data = colorectal, recurrentAG = TRUE,
+    n.knots = 8, cross.validation = T, kappa = 1000, RandDist = "LogN")
\end{CodeInput}
\end{CodeChunk}
Then, using argument \code{init.B} we fitted the trivariate model with appropriate initial values. In \code{init.B}, the vector of initial values must follow the order: covariates for the recurrent events, terminal event and then  biomarker (interactions in the end of each component). The trivariate model due to its complexity is computationally intensive but using the pseudo-adaptive quadrature method the time of estimation is reduced (40 minutes on a personal computer with an Intel Core i7 3.40 GHz processor and 8 GB RAM runinng Windows 7). The model is defined as follows:
\begin{CodeChunk}
\begin{CodeInput}
R> modTrivariate <- trivPenal(Surv(time0, time1, new.lesions) ~ cluster(id) + 
+    age + treatment + who.PS + terminal(state), formula.terminalEvent = ~ 
+    age + treatment + who.PS + prev.resection, tumor.size ~ year * 
+    treatment + age + who.PS, data = colorectal, data.Longi = 
+    colorectalLongi, random = c("1", "year"), id = "id", link = 
+    "Random-effects", left.censoring = -3.33, recurrentAG = T, n.knots = 7,
+    kappa = c(0.01, 0.7), method.GH = "Pseudo-adaptive", n.nodes = 7,
+    init.B = c(-0.18, -0.22, -0.24, -0.22, 0.35, -0.23, -0.10, -0.09, -0.12,
+    0.80, -0.23, 3.02, -0.30, 0.05, -0.02, -0.29, 0.11, 0.74, -0.63))
R> modTrivariate
\end{CodeInput}
\end{CodeChunk}
\begin{CodeChunk}
\begin{CodeOutput}
Call:
trivPenal(formula = Surv(time0, time1, new.lesions) ~ cluster(id) + 
    age + treatment + who.PS + terminal(state), formula.terminalEvent = ~age + 
    treatment + who.PS + prev.resection, formula.LongitudinalData = tumor.size ~ 
    year * treatment + age + who.PS, data = colorectal, data.Longi = colorectalLongi, 
    random = c("1", "year"), id = "id", link = "Random-effects", 
    left.censoring = -3.33, recurrentAG = TRUE, n.knots = 7, 
    kappa = c(0.01, 0.7), init.B = c(-0.18, -0.22, -0.24, -0.22, 
    0.35, -0.23, -0.1, -0.09, -0.12, 0.8, -0.23, 3.02, -0.3, 
    0.05, -0.02, -0.29, 0.11, 0.74, -0.63), method.GH = "Pseudo-adaptive", 
    n.nodes = 7)

      Calendar timescale

   Trivariate Joint Model for Longitudinal Data, Recurrent Events and a Terminal Event 
   Parameter estimates using a Penalized Likelihood on the hazard functions 
   and assuming left-censored longitudinal outcome 
   Association function: random effects 


Longitudinal outcome:
---------------- 
                     coef SE coef (H) SE coef (HIH)         z          p
Intercept        2.958825    0.187712      0.186829 15.762577     <1e-16
year            -0.282875    0.134839      0.134696 -2.097872 3.5916e-02
treatmentC       0.102905    0.215246      0.214927  0.478081 6.3259e-01
age60-69 years   0.013937    0.167633      0.167385  0.083138 9.3374e-01
age>69 years    -0.272289    0.131499      0.131356 -2.070649 3.8392e-02
who.PS1          0.120884    0.116501      0.116448  1.037616 2.9945e-01
who.PS2          0.760509    0.175385      0.175368  4.336218 1.4496e-05
year:treatmentC -0.664342    0.185324      0.185216 -3.584762 3.3739e-04

          chisq df global p
age     6.50037  2 0.038800
who.PS 19.25190  2 0.000066

Recurrences:
------------- 
                    coef exp(coef) SE coef (H) SE coef (HIH)         z 
age60-69 years -0.272583  0.761410    0.272273      0.268259 -1.001141 
age>69 years   -0.298342  0.742047    0.250754      0.249399 -1.189780 
treatmentC     -0.290463  0.747917    0.219911      0.216209 -1.320820 
who.PS1        -0.008039  0.991993    0.263756      0.257795 -0.030479 
who.PS2         0.691122  1.995953    0.284057      0.280034  2.433035 

         p
3.1676e-01 
2.3413e-01   
1.8656e-01
9.7568e-01
1.4973e-02

         chisq df global p
age    1.74005  2    0.419
who.PS 6.70573  2    0.035

Terminal event:
---------------- 
                       coef exp(coef) SE coef (H) SE coef (HIH)         z 
age60-69 years    -0.250844  0.778143    0.543595      0.522310 -0.461454 
age>69 years      -0.319804  0.726291    0.467831      0.461083 -0.683589 
treatmentC        -0.103662  0.901530    0.409085      0.389645 -0.253401 
who.PS1            0.521433  1.684439    0.598905      0.575073  0.870643 
who.PS2            2.155355  8.630949    0.633033      0.609088  3.404809 
prev.resectionYes -0.071156  0.931317    0.390449      0.379241 -0.182241 

         p
6.4447e-01
4.9423e-01
7.9996e-01
3.8395e-01
6.6210e-04
8.5539e-01

           chisq df global p
age    10.810677  2  0.00449
who.PS  0.265533  2  0.87600

 
Components of Random-effects covariance matrix B1: 
                             
Intercept  1.983237 -0.526571
year      -0.526571  0.955922

Recurrent event and longitudinal outcome association: 
                     coef        SE         z       p
Asso:Intercept  0.1368545 0.0965219  1.417860 0.15623
Asso:year      -0.0674389 0.1340421 -0.503117 0.61488

Terminal event and longitudinal outcome association: 
                     coef       SE        z         p
Asso:Intercept  0.8456482 0.290082  2.91520 0.0035546
Asso:year      -0.0575677 0.234722 -0.24526 0.8062600

Residual standard error:  0.13981  (SE (H):  0.234722 ) 
 
 Frailty parameter for the association between recurrent events and terminal event: 
   sigma square (variance of Frailties): 0.555944 (SE (H): 0.406022 ) p = 0.085461 
   alpha (for terminal event): 2.624 (SE (H): 0.250754 ) p = <1e-16 
 
   penalized marginal log-likelihood = -1929.42
   Convergence criteria: 
   parameters = 0.00016 likelihood = 0.000926 gradient = 9.45e-06 

   LCV = the approximate likelihood cross-validation criterion
         in the semi parametric case     = 1.64853 

   n subjects= 150
   n repeated measurements= 906
      Percentage of left-censored measurements= 3.75 %
      Censoring threshold s= -3.33

   n recurrent events= 139
   n terminal events= 121

   number of iterations:  24 

   Exact number of knots used:  7 
   Value of the smoothing parameters:  0.01 0.7

      Gaussian quadrature method:  Pseudo-adaptive with 7 nodes 
\end{CodeOutput}
\end{CodeChunk}

In the output of the fitted model, we obtain the estimates for the covariates related to the three types of processes: longitudinal outcome, recurrences and terminal event. The treatment effect was not significant neither for the risk of appearance of new lesions nor for the risk of death. As in the bivariate model, the tumor size decreased on average more in the treatment arm C ($-0.66$, p<$0.001$). We found the baseline WHO performance status 2 having a significant effect on the average population biomarker, the risk of new lesions appearance and death. The coefficients for the association between the biomarker and the recurrences were found not significant and for the link between the biomarker and death only the coefficient related to the random intercept was significantly different from 0 ($0.85$, p=$0.004$). The variance of the frailty term was found to be not significant.

As for the bivariate model \code{modLongi}, goodnes-of-fit of the trivariate model can be evaluated using the martingale residuals for the recurrent and terminal events and the residuals related to the biomarker. For this purpose, a following code can be used (Figure~\ref{figure:residuals_triv}, Appendix B):
\begin{CodeChunk}
\begin{CodeInput}
R> plot(modTrivariate, main = "Hazard functions")
R> plot(modTrivariate, type = "Survival", main = "Survival functions")
R> plot(aggregate(colorectal$time1, by = list(colorectal$id), FUN = 
+    max)[2][ ,1], modTrivariate$martingaledeath.res, ylab = "", xlab = 
+    "time", main = "Martingale Residuals - Death", ylim = c(-4.2, 4.2))
R> lines(lowess(aggregate(colorectal$time1, by = list(colorectal$id),
+    FUN = max)[2][ ,1], modTrivariate$martingaledeath.res, f = 1), lwd = 3, 
+    col = "grey")
R> plot(aggregate(colorectal$time1, by = list(colorectal$id), FUN = 
+    max)[2][ ,1], modTrivariate$martingale.res, ylab = "", xlab = "time",     
+    main = "Martingale Residuals\n  - Recurrences", ylim = c(-4.2, 4.2))
R> lines(lowess(aggregate(colorectal$time1, by = list(colorectal$id),
+    FUN = max)[2][ ,1], modTrivariate$martingale.res, f = 1), lwd = 3,
+    col = "grey")
R> qqnorm(modTrivariate$marginal_chol.res, main = "Marginal Cholesky 
+    residuals" ,xlab = "")
R> qqline(modTrivariate$marginal_chol.res)
R> plot(modTrivariate$pred.y.cond, modTrivariate$conditional.res, 
+    xlab = "Fitted", ylab = "Conditional residuals", main = "Conditional 
+    Residuals \n vs. Fitted Values")
\end{CodeInput}
\end{CodeChunk}
The martingale residuals for both recurrences and death processes showed slight signs of skewness, their means seemed to decrease in time. Using the marginal Cholesky residuals and the conditional residuals we found that the model did not fit the longitudinal data very well.

We compared the trivariate model \code{modTrivariate} with the bivariate model \code{modLongi} in terms of predictive accuracy for OS using the Brier Score. Firstly, we defined the prediction time and horizons and calculated the predictions for the bivariate and trivariate models: 
\begin{CodeChunk}
\begin{CodeInput}
R> predtime <- 1.0
R> window <- seq(0.1, 1.5, 0.1)
R> fwindow <- predtime + window
R> pred_bivariate<- prediction(modLongi, data = colorectalSurv, 
+    data.Longi = colorectalLongi, predtime, window)
R> pred_trivariate <- prediction(modTrivariate, data = colorectal, 
+    data.Longi = colorectalLongi, predtime, window)
\end{CodeInput}
\end{CodeChunk}
We prepared data for the Brier Score computations: \code{data_surv} including data on survival (variables of time and state), \code{survj_biv} and \code{survj_tri} for predicted probabilities of survival of subjects that were alive at time of prediction using the bivariate and trivariate model, respectively:
\begin{CodeChunk}
\begin{CodeInput}
R> data_surv <- colorectalSurv[ , c(1, 3, 9)]
R> predictions_bivariate <- as.matrix(pred_bivariate$pred)
R> survj_biv <- as.matrix(predictions_bivariate[data_surv$time1 >= predtime, ])
R> survj_biv <- 1 - cbind(0, survj_biv)
R> predictions_trivariate <- as.matrix(pred_trivariate$pred)
R> survj_tri <- as.matrix(predictions_trivariate[data_surv$time1 >=
+    predtime, ])
R> survj_tri <- 1 - cbind(0, survj_tri)
\end{CodeInput}
\end{CodeChunk}
Using adjusted functions from the package \pkg{pec} (available on request from the authors), we obtained the apparent prediction error  curves (using the data used for the estimation)as follows:
\begin{CodeChunk}
\begin{CodeInput}
R> library("pec")
R> library("prodlim")
R> source("/ipcw_2-9_modif.R")
R> source("/pecMethods_2-9_modif.R")
R> BrierScore <- pec(list("Bivariate" = survj_biv, "Trivariate" = survj_tri),
+    formula = Surv(time1, state) ~ 1, data = data_surv[data_surv$time1 
+    >= predtime, ], cens.model = "marginal", data.cens = data_surv, 
+    exact = FALSE, times = fwindow, ptime = predtime, reference = FALSE)
R> plot(fwindow, BrierScore$AppErr$Bivariate[-1], pch = 6, main = "", 
+    ylab = "Prediction error", xlab = "Years", ylim = c(0, 0.3), xlim = 
+    c(0.9, 2.7), col = "black", type = "l", lwd = 2, axes = F, lty = 1)
R> points(fwindow, BrierScore$AppErr$Trivairate[-1], pch = 15, col = "blue",
+    type = "l", lwd = 2, lty = 2)
R> legend(2.1, 0.3, legend = c("modLongi", "modTrivariate"), bty = "n", 
+    cex = 1.1, lty = c(1, 2), lwd = 2, col = c("black", "blue"))
R> axis(1, at = seq(0.9, 2.7, by = 0.3))
R> axis(2, at = seq(0, 0.3, by = 0.1))
R> abline(v = 1, lty = 2, lwd = 2, col = "gray35")
\end{CodeInput}
\end{CodeChunk}

\begin{figure}[h!]
\begin{center}
\includegraphics[ trim = 0cm 0.5cm 0.5cm 2cm, clip]{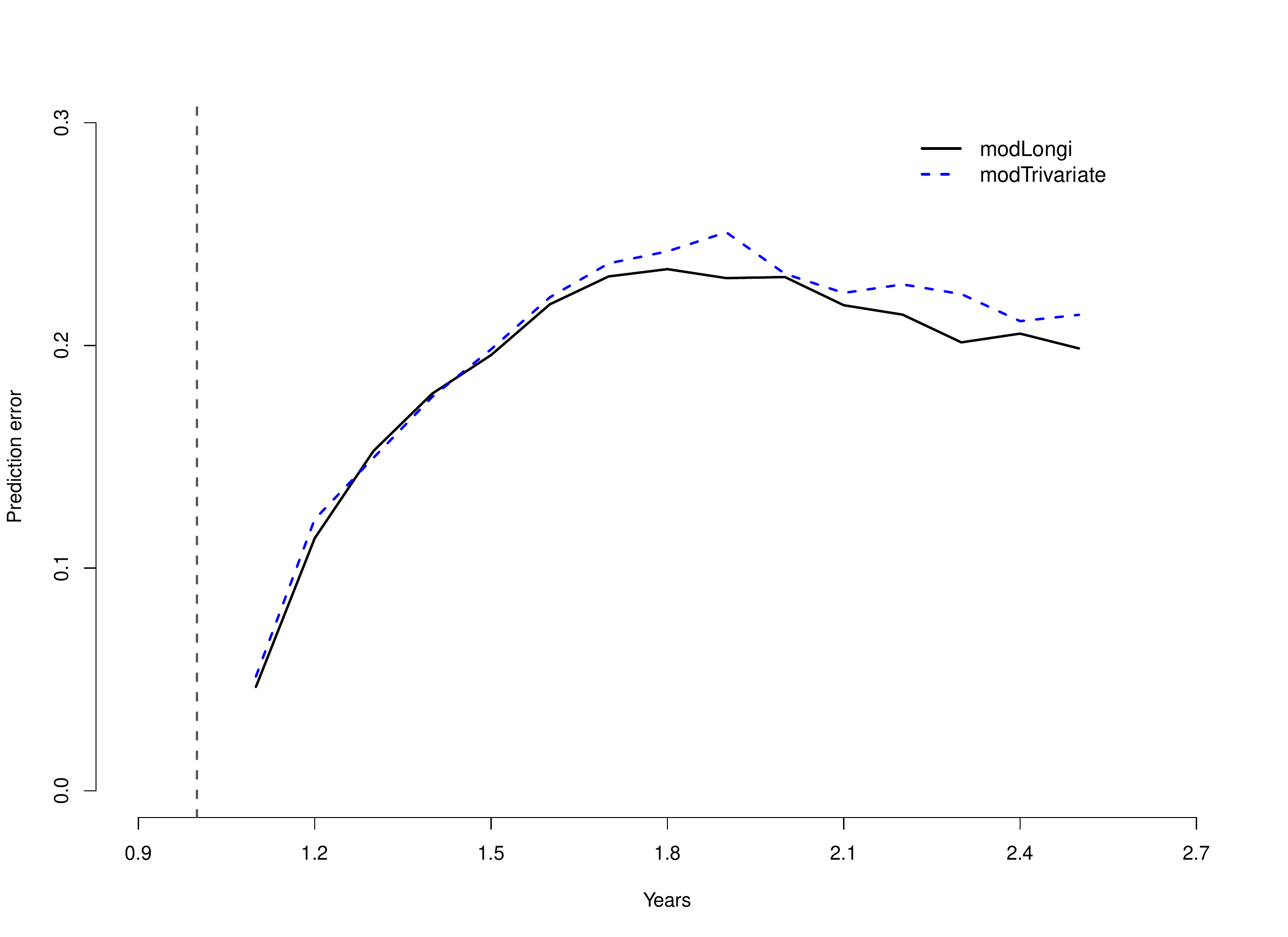} 
	\caption{Error of prediction at t = 1 year and varying window $w$
from 0.1 to 1.5. \code{modLongi} -  tumor size and death, \code{modTrivariate} - tumor size, occurrence of new
lesions and death.}
	\label{figure:BrierScore}
	\end{center}
\end{figure}

Figure~\ref{figure:BrierScore} represents the prediction error curves for both models. In short  prediction horizons, values of the Brier Score of the models were very close to each other. After around 1.6 years the bivariate model had better predictive accuracy than the trivariate model. However, the results are presented for the apparent error calculated using the data from the estimation. For verification of the models for predictions for new patients (not used in the data for the estimation) a cross-validation procedure would be required.

\subsubsection{Dynamic Predictions}

Both, for the bivariate models for longitudinal data and a terminal event and for the trivariate models, the package provides individual predictions of the terminal event. To create patients' profiles, two datasets must be provided, one including the history of the biomarker and the values of covariates at measurement times (\code{dataLongi}) and a second one with the information on covariates related to the terminal event (\code{data}). In case of the trivariate model, this dataset should also include the history of recurrences. 

For the example of \code{colorectal} dataset, we created two profiles of patients that differed from each other by the trajectory of the tumor size. The first patient had a progressive disease with a tumor size increasing in time and the second patient a response to the treatment with a diminishing tumor. Both patients had the same baseline characteristics. In the step of creating the data, types of variables must be appropriate and coherent with the dataset used for estimation. Firstly, we prepared the data for the bivariate model setting:

\begin{CodeChunk}
\begin{CodeInput}
R> datapredj_longi <- data.frame(id = 0, year = 0, tumor.size = 0, 
+    treatment = 0, age = 0, who.PS = 0, prev.resection = 0)
R> datapredj_longi$treatment <- factor(datapredj_longi$treatment, 
+    levels = 1 : 2)
R> datapredj_longi$age <- factor(datapredj_longi$age, levels = 1 : 3)
R> datapredj_longi$who.PS <- factor(datapredj_longi$who.PS, levels = 1 : 3)
R> datapredj_longi$prev.resection <- factor(datapredj_longi$prev.resection, 
+    levels = 1 : 2)
\end{CodeInput}
\end{CodeChunk}
For Patient 1 we assumed 5 measurements indicating increasing tumor burden:
\begin{CodeChunk}
\begin{CodeInput}
R> datapredj_longi[1, ] <- c(1, 0, 1.2, 2, 1, 1, 1)
R> datapredj_longi[2, ] <- c(1, 0.3, 1.1, 2, 1, 1, 1)
R> datapredj_longi[3, ] <- c(1, 0.6, 1.4, 2, 1, 1, 1)
R> datapredj_longi[4, ] <- c(1, 0.9, 2.2, 2, 1, 1, 1)
R> datapredj_longi[5, ] <- c(1, 1.5, 3.0, 2, 1, 1, 1)
\end{CodeInput}
\end{CodeChunk}
On the contrary, Patient 2 was assumed to have a decreasing size of the tumor:
\begin{CodeChunk}
\begin{CodeInput}
R> datapredj_longi[6,] <- c(2, 0, 1.2, 2, 1, 1, 1)
R> datapredj_longi[7,] <- c(2, 0.3, 1.1, 2, 1, 1, 1)
R> datapredj_longi[8,] <- c(2, 0.5, 0.7, 2, 1, 1, 1)
R> datapredj_longi[9,] <- c(2, 0.7, 0.3, 2, 1, 1, 1)
R> datapredj_longi[10,] <- c(2, 0.9, 0.1, 2, 1, 1, 1)
\end{CodeInput}
\end{CodeChunk}
Next, for both patients we prepared the data with information on covariates included in the survival part in the bivariate model:
\begin{CodeChunk}
\begin{CodeInput}
R> datapredj <- data.frame(id = 0, treatment = 0, age = 0, who.PS = 0, 
+    prev.resection = 0)
R> datapredj$treatment <- factor(datapredj$treatment, levels = 1 : 2)
R> datapredj$age <- factor(datapredj$age, levels = 1 : 3)
R> datapredj$who.PS <- factor(datapredj$who.PS, levels = 1 : 3)
R> datapredj$prev.resection <- factor(datapredj$prev.resection, levels = 1 : 2)
R> datapredj[1, ] <- c(1, 2, 1, 1, 1)
R> datapredj[2, ] <- c(2, 2, 1, 1, 1)
\end{CodeInput}
\end{CodeChunk}

We calculated the estimated probabilities of the terminal event given that the patients were alive at a time of prediction 1 year and a horizon varying from 0.5 to 2.5 years. We compared the predicted risk of death for the patients by plotting and smoothing the estimations. Additionally, the 95\% MC confidence intervals were calculated in order to facilitate the interpretation.
\begin{CodeChunk}
\begin{CodeInput}
R> pred.joint <- prediction(modLongi, datapredj, datapredj_longi, 1.0, 
+    seq(0.5, 2.5, 0.2), MC.sample = 500)
R> plot(pred.joint, conf.bands = TRUE)
\end{CodeInput}
\end{CodeChunk}

\begin{figure}[h!]
\begin{center}
	\includegraphics[ trim = 0.0cm 0.6cm 0.2cm 0.2cm, clip ]{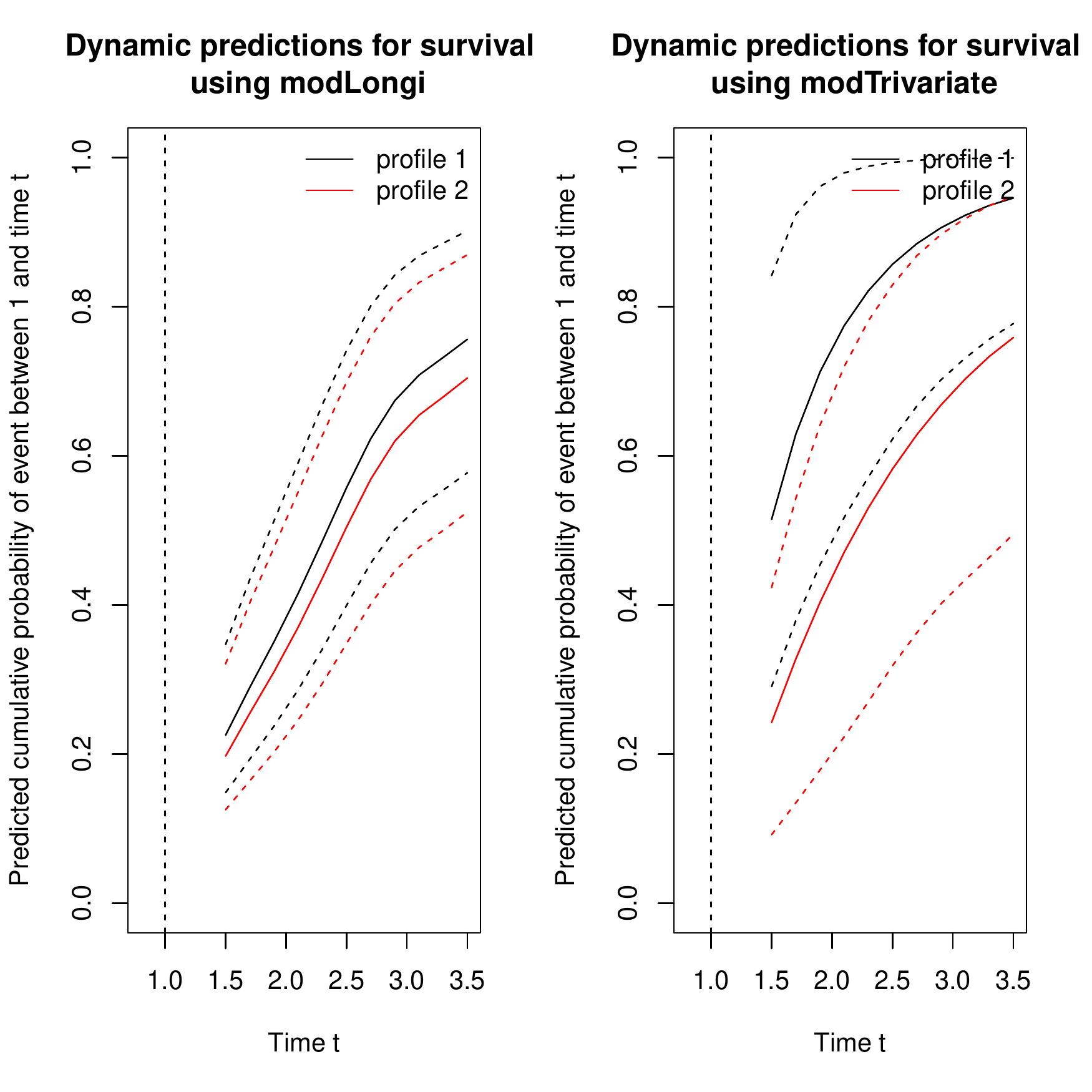} 
	\caption{Predicted probability of death for two patients sharing the same characteristics and different history using information from the biomarker only (on the left) and the biomarker and recurrent events (on the right).  The prediction time was set to 1 year and the horizon changed from 0.5 to 2.5 years. Dashed lines represent the MC confidence intervals.}
	\label{figure:prediction2}
	\end{center}
\end{figure}

The left graph in Figure~\ref{figure:prediction2} present the dynamic predictions for the patients. The patient with a decreasing tumor size (profile 2) had lower probability of death than the patient with the tumor size that increased during the treatment (profile 1). However, considering the MC confidence intervals, this difference was not significant. Thus, in the analyzed example, the biomarker itself does not influence the risk of death significantly. It is of interest if  addition of the history of recurrent event would increase the difference between the profiles. For this purpose, we modified data \code{datapredj} by adding the history of recurrences and calculated the analogous dynamic predictions using trivariate model \code{modTrivariate}. We assumed that patient 1 experienced the occurrence of new lesions twice and patient 2 only once.

\begin{CodeChunk}
\begin{CodeInput}
R> datapredj <- data.frame(time0 = 0, time1 = 0, new.lesions = 0, id = 0, 
+    treatment = 0, age = 0, who.PS = 0, prev.resection = 0)
R> datapredj$treatment <- factor(datapredj$treatment, levels = 1 : 2)
R> datapredj$age <- factor(datapredj$age, levels = 1 : 3)
R> datapredj$who.PS <- factor(datapredj$who.PS, levels = 1 : 3)
R> datapredj$prev.resection <- factor(datapredj$prev.resection, levels = 1 : 2)
R> datapredj[1, ] <- c(0, 0.4, 1, 1, 2, 1, 1, 1)
R> datapredj[2, ] <- c(0.4, 1.2, 1, 1, 2, 1, 1, 1)
R> datapredj[3, ] <- c(0, 0.5, 1, 2, 2, 1, 1, 1)
\end{CodeInput}
\end{CodeChunk}
Then, we calculated the predictions and plotted the results with the 95\% MC confidence intervals:
\begin{CodeChunk}
\begin{CodeInput}
R> pred.joint2 <- prediction(modTrivariate, datapredj, datapredj_longi, 1.0,
+    seq(0.5, 2.5, 0.2), MC.sample = 500)
R> plot(pred.joint2, conf.bands = TRUE)
\end{CodeInput}
\end{CodeChunk}

The right graph in Figure~\ref{figure:prediction2} shows the dynamic predictions of death using the trivariate model. As in the case of the bivariate model, patient 1 had an increased probability of death comparing to patient 1. Moreover, the difference between the patients was more accentuated than in the bivariate model that was not able to include information on the history of recurrences. By considering the information on the appearance of new lesions, the predicted probabilities were influence, although the differences between the patients were not significant according to the MC confidence intervals.

		\section{Conclusions}
		\label{conclusion}
Joint models are now a well recognized statistical tool for a complex analysis of correlated data. They provide unbiased estimates comparing to the univariate models. The package \pkg{frailtypack} provides several functions for applications of the joint models for a longitudinal outcome, recurrent events and a terminal event. Methods of estimation applied in the package produce reliable results proven by simulations for all the implemented models (\citet{Rondeau2007joint}, \citet{Mazroui2012}, \citet{Krol2015}). Moreover, each function was furnished with several options for better adjustment of a model and in tools helping the diagnostic of a model (residuals, individual predictions of random effects) and comparison with other models (LCV, AIC). 

The increasing interest in individual predictions of death in the clinical perspective motivated the implementation of model-based dynamic predictions. Function \code{prediction} allows the users to calculate the estimated probability of death given the history of a patient. The history, depending on a model, can be a part of information on observed recurrences, complete information on observed recurrences, complete past measurements of a biomarker or the entire available history of the recurrences and biomarker. Graphical representation of the predictions is a useful tool for the comparing probability of death of patients e.g., with the same characteristics but different past of recurrences and/or biomarker. 

Further developments of the \pkg{frailtypack} will concern the extensions of the existent functions. These extensions will be related to the random effects (distributions, number of random effects for the longitudinal part), association functions for models with a biomarker (provide more forms to choose from), stratification (increase the number of strata) and dynamic predictions for the joint model with two types of recurrent events and a terminal event. Moreover, in order to increase the variety of possible application, it will be of interest to implement the left truncation and the interval-censoring in the proposed functions. 

In the near future, we will develop dynamic predictions of a recurrent event. The conditional probability of occurring a new recurrence in a finite time horizon given the history of the observed events and/or the biomarker and given that an individual is alive at the prediction time, will constitute a useful tool for clinicians e.g., to explore the chances of developing a new relapse given a patient's characteristics.

\section*{Acknowledgments}
The authors thank the F\'{e}d\'{e}ration Francophone de Canc\'{e}rologie Digestive and Gustave Roussy for sharing the data of the
FFCD 2000-05 trial supported by an unrestricted Grant from Sanofi. The authors thank as well the IReSP for financing the project AAP 2012 Plan Cancer MULTIPLE, the Institut National du Cancer for funding the research of A. Mauguen the \'{E}cole des Hautes \'{E}tudes en Sant\'{e} Publique for funding the research of A. Kr\'{o}l.  

\bibliography{lib_jss} 

\begin{thebibliography}{56}
\newcommand{\enquote}[1]{``#1''}
\providecommand{\natexlab}[1]{#1}
\providecommand{\url}[1]{\texttt{#1}}
\providecommand{\urlprefix}{URL }
\expandafter\ifx\csname urlstyle\endcsname\relax
  \providecommand{\doi}[1]{doi:\discretionary{}{}{}#1}\else
  \providecommand{\doi}{doi:\discretionary{}{}{}\begingroup
  \urlstyle{rm}\Url}\fi
\providecommand{\eprint}[2][]{\url{#2}}

\bibitem[{Andersen and Gill(1982)}]{Andersen1982}
Andersen PK, Gill RD (1982).
\newblock \enquote{{Cox's Regression Model for Counting Processes: A Large
  Sample Study}.}
\newblock \emph{The Annals of Applied Statistics}, pp. 1100--1120.

\bibitem[{Asar \emph{et~al.}(2015)Asar, Ritchie, Kalra, and Diggle}]{Asar2015}
Asar O, Ritchie J, Kalra PA, Diggle P (2015).
\newblock \enquote{{Joint Modelling of Repeated Measurement and Time-to-Event
  Data: An Introductory Tutorial}.}
\newblock \emph{International Journal of Epidemiology}, \textbf{44}(1),
  334--344.

\bibitem[{Belot \emph{et~al.}(2014)Belot, Rondeau, Remontet, Roch, and {CENSUR
  working survival group}}]{Belot2014}
Belot A, Rondeau V, Remontet L, Roch G, {CENSUR working survival group} (2014).
\newblock \enquote{{A Joint Frailty Model to Estimate the Recurrence Process
  and the Disease-Specific Mortality Process without Needing the Cause of
  Death}.}
\newblock \emph{Statistics in Medicine}, \textbf{33}, 3147--3166.

\bibitem[{Blanche \emph{et~al.}(2015)Blanche, Proust-Lima, Loub\`{e}re, Berr,
  Dartigues, and Jacqmin-Gadda}]{Blanche2015}
Blanche P, Proust-Lima C, Loub\`{e}re L, Berr C, Dartigues JF, Jacqmin-Gadda H
  (2015).
\newblock \enquote{{Quantifying and Comparing Dynamic Predictive Accuracy of
  Joint Models for Longitudinal Marker and Time-to-Event in Presence of
  Censoring and Competing Risks}.}
\newblock \emph{Biometrics}, \textbf{71}, 102--113.

\bibitem[{Chi and Ibrahim(2006)}]{Chi2006}
Chi YY, Ibrahim JG (2006).
\newblock \enquote{{Joint Models for Multivariate Longitudinal and Multivariate
  Survival Data}.}
\newblock \emph{Biometrics}, \textbf{62}, 432--445.

\bibitem[{Commenges \emph{et~al.}(2012)Commenges, Liquet, and
  Proust-Lima}]{Commenges2012}
Commenges D, Liquet B, Proust-Lima C (2012).
\newblock \enquote{{Choice of Prognostic Estimators in Joint Models by
  Estimating Differences of Expected Conditional Kullback-Leibler Risks.}}
\newblock \emph{Biometrics}, \textbf{68}(2), 380--7.

\bibitem[{Commenges and Rondeau(2000)}]{Commenges2000}
Commenges D, Rondeau V (2000).
\newblock \enquote{{Standardized Martingale Residuals Applied to Grouped Left
  Truncated Observations of Dementia Cases}.}
\newblock \emph{Lifetime Data Analysis}, \textbf{6}, 229--235.

\bibitem[{Crowther(2013)}]{stjm}
Crowther MJ (2013).
\newblock \emph{\pkg{stjm}: \proglang{Stata} Mmodule to Fit Shared Parameter
  Joint Models of Longitudinal and Survival Data}.
\newblock
  \urlprefix\url{http://econpapers.repec.org/software/bocbocode/s457502.htm}.

\bibitem[{{De Boor}(2001)}]{DeBoor2001}
{De Boor} C (2001).
\newblock \emph{{A Practical Guide to Splines}}.
\newblock Springer-Verlag, Berlin.

\bibitem[{Duchateau \emph{et~al.}(2003)Duchateau, Janssen, Kezic, and
  Fortpied}]{Duchateau2003}
Duchateau L, Janssen P, Kezic I, Fortpied C (2003).
\newblock \enquote{{Evolution of Recurrent Asthma Event Rate over Time}.}
\newblock \emph{Journal of the Royal Statistical Society C (Applied
  Statistics)}, \textbf{52}(3), 355--363.

\bibitem[{Ducreux \emph{et~al.}(2011)Ducreux, Malka, Mendiboure, Etienne,
  Texereau, Auby, Rougier, Gasmi, Castaing, Abbas, Michel, Gargot, Azzedine,
  Lombard-Bohas, Geoffroy, Denis, Pignon, Bedenne, and
  Bouch\'{e}}]{Ducreux2011}
Ducreux M, Malka D, Mendiboure J, Etienne PL, Texereau P, Auby D, Rougier P,
  Gasmi M, Castaing M, Abbas M, Michel P, Gargot D, Azzedine A, Lombard-Bohas
  C, Geoffroy P, Denis B, Pignon JP, Bedenne L, Bouch\'{e} O (2011).
\newblock \enquote{{Sequential versus Combination Chemotherapy for the
  Treatment of Advanced Colorectal Cancer (FFCD 2000-05): An Open-Label,
  Randomised, Phase 3 Trial}.}
\newblock \emph{The Lancet Oncology}, \textbf{12}(11), 1032--44.

\bibitem[{Efendi \emph{et~al.}(2013)Efendi, Molenberghs, Njagi, and
  Dendale}]{Efendi2013}
Efendi A, Molenberghs G, Njagi EN, Dendale P (2013).
\newblock \enquote{{A Joint Model for Longitudinal Continuous and Time-to-Event
  Outcomes with Direct Marginal Interpretation}.}
\newblock \emph{Biometrical Journal}, \textbf{55}(4), 572--88.

\bibitem[{Elashoff \emph{et~al.}(2008)Elashoff, Li, and Li}]{Elashoff2008}
Elashoff RM, Li G, Li N (2008).
\newblock \enquote{{A Joint Model for Longitudinal Measurements and Survival
  Data in the Presence of Multiple Failure Types}.}
\newblock \emph{Biometrics}, \textbf{64}(3), 762--771.

\bibitem[{Emura(2016)}]{joint.Cox}
Emura T (2016).
\newblock \emph{\pkg{joint.Cox}: Penalized Likelihood Estimation under the
  Joint Cox Models Between TTP and OS for Meta-Analysis}.
\newblock \proglang{R}~package version~2.6,
  \urlprefix\url{http://CRAN.R-project.org/package=joint.Cox}.

\bibitem[{Genz and Keister(1996)}]{Genz1996}
Genz A, Keister BD (1996).
\newblock \enquote{{Fully Symmetric Interpolatory Rules for Multiple Integrals
  over Infinite Regions with Gaussian Weight}.}
\newblock \emph{Journal of Computational and Applied Mathematics},
  \textbf{71}(2), 299--309.

\bibitem[{Gerds(2016)}]{pec}
Gerds TA (2016).
\newblock \emph{\pkg{pec}: Prediction Error Curves for Risk Prediction Models
  in Survival Analysis}.
\newblock \proglang{R}~package version~2.4.9,
  \urlprefix\url{http://CRAN.R-project.org/package=pec}.

\bibitem[{Gerds and Schumacher(2006)}]{Gerds2006}
Gerds TA, Schumacher M (2006).
\newblock \enquote{{Consistent Estimation of the Expected Brier Score in
  General Survival Models with Right-Censored Event Times}.}
\newblock \emph{Biometrical Journal}, \textbf{48}(6), 1029--1040.

\bibitem[{Gonzalez \emph{et~al.}(2005)Gonzalez, Fernandez, Moreno, Ribes,
  Merce, Navarro, Cambray, and Borras}]{Gonzalez2005}
Gonzalez JR, Fernandez E, Moreno V, Ribes J, Merce P, Navarro M, Cambray M,
  Borras JM (2005).
\newblock \enquote{{Sex Differences in Hospital Readmission among Colorectal
  Cancer Patients}.}
\newblock \emph{Journal of Epidemiology and Community Health}, \textbf{59}(6),
  506--511.

\bibitem[{Hatfield \emph{et~al.}(2012)Hatfield, Boye, Hackshaw, and
  Carlin}]{Hatfield2012}
Hatfield LA, Boye ME, Hackshaw MD, Carlin BP (2012).
\newblock \enquote{{Multilevel Bayesian Models for Survival Times and
  Longitudinal Patient-Reported Outcomes with Many Zeros}.}
\newblock \emph{Journal of the American Statistical Association},
  \textbf{107}(499), 875--885.

\bibitem[{Joly \emph{et~al.}(1998)Joly, Commenges, and Lettenneur}]{Joly1998}
Joly P, Commenges D, Lettenneur L (1998).
\newblock \enquote{{A Penalized Likelihood Approach for Arbitrarily Censored
  and Truncated Data: Application to Age-Specific Incidence of Dementia}.}
\newblock \emph{Biometrics}, \textbf{54}(1), 185--194.

\bibitem[{Kim \emph{et~al.}(2012)Kim, Zeng, Chambless, and Li}]{Kim2012}
Kim S, Zeng D, Chambless L, Li Y (2012).
\newblock \enquote{{Joint Models of Longitudinal Data and Recurrent Events with
  Informative Terminal Event}.}
\newblock \emph{Statistics in Bioscience}, \textbf{4}(2), 262--281.

\bibitem[{Kr\'{o}l \emph{et~al.}(2016)Kr\'{o}l, Ferrer, Pignon, Proust-Lima,
  Ducreux, Bouch\'{e}, Michiels, and Rondeau}]{Krol2015}
Kr\'{o}l A, Ferrer L, Pignon JP, Proust-Lima C, Ducreux M, Bouch\'{e} O,
  Michiels S, Rondeau V (2016).
\newblock \enquote{{Joint Model for Left-Censored Longitudinal Data, Recurrent
  Events and Terminal Event: Predictive Abilities of Tumor Burden for Cancer
  Evolution with Application to the FFCD 2000-05 Trial}.}
\newblock \emph{Biometrics}.

\bibitem[{{Lawrence Gould} \emph{et~al.}(2014){Lawrence Gould}, Boye, Crowther,
  Ibrahim, Quartey, Micallef, and Bois}]{LawrenceGould2014}
{Lawrence Gould} A, Boye ME, Crowther MJ, Ibrahim JG, Quartey G, Micallef S,
  Bois FY (2014).
\newblock \enquote{{Joint Modeling of Survival and Longitudinal Non-Survival
  Data: Current Methods and Issues. Report of the DIA Bayesian Joint Modeling
  Working Group}.}
\newblock \emph{Statistics in Medicine}.

\bibitem[{Li and Lagakos(1997)}]{Li1997}
Li QH, Lagakos SW (1997).
\newblock \enquote{{Use of the Wei-Lin-Weissfeld method for the analysis of a
  recurring and a terminating event}.}
\newblock \emph{Statistics in Medicine}, \textbf{16}, 925--940.

\bibitem[{Liu and Huang(2009)}]{Liu2009a}
Liu L, Huang X (2009).
\newblock \enquote{{Joint Analysis of Correlated Repeated Measures and
  Recurrent Events Processes in the Presence of Death, with Application to a
  Study on Acquired Immune Deficiency Syndrome}.}
\newblock \emph{Journal of the Royal Statistical Society}, \textbf{58}(1),
  65--81.

\bibitem[{Liu \emph{et~al.}(2008)Liu, Huang, and O'Quigley}]{Liu2008}
Liu L, Huang X, O'Quigley JM (2008).
\newblock \enquote{{Analysis of Longitudinal Data in the Presence of
  Informative Observational Times and a Dependent Terminal Event, with
  Application to Medical Cost Data.}}
\newblock \emph{Biometrics}, \textbf{64}(3), 950--8.

\bibitem[{Liu \emph{et~al.}(2004)Liu, Wolfe, and Huang}]{Liu2004}
Liu L, Wolfe RA, Huang X (2004).
\newblock \enquote{{Shared Frailty Models for Recurrent Events and a Terminal
  Event.}}
\newblock \emph{Biometrics}, \textbf{60}(3), 747--56.

\bibitem[{Marquardt(1963)}]{Marquardt1963}
Marquardt DW (1963).
\newblock \enquote{{An Algorithm for Least-Squares Estimation of Nonlinear
  Parameters}.}
\newblock \emph{Journal of the Society for Industrial and Applied Mathematics},
  \textbf{11}(2), 431--41.

\bibitem[{Mauguen \emph{et~al.}(2013)Mauguen, Rachet, Mathoulin-P\'{e}lissier,
  MacGrogan, Laurent, and Rondeau}]{Mauguen2013}
Mauguen A, Rachet B, Mathoulin-P\'{e}lissier S, MacGrogan G, Laurent A, Rondeau
  V (2013).
\newblock \enquote{{Dynamic Prediction of Risk of Death Using History of Cancer
  Recurrences in Joint Frailty Models.}}
\newblock \emph{Statistics in Medicine}, \textbf{32}(30), 5366--80.

\bibitem[{Mazroui \emph{et~al.}(2013)Mazroui, Mathoulin-P\'{e}lissier,
  Macgrogan, Brouste, and Rondeau}]{Mazroui2013}
Mazroui Y, Mathoulin-P\'{e}lissier S, Macgrogan G, Brouste V, Rondeau V (2013).
\newblock \enquote{{Multivariate Frailty Models for Two Types of Recurrent
  Events with a Dependent Terminal Event: Application to Breast Cancer Data.}}
\newblock \emph{Biometrical Journal}, \textbf{55}(6), 866--84.

\bibitem[{Mazroui \emph{et~al.}(2012)Mazroui, Mathoulin-P\'{e}lissier,
  Soubeyran, and Rondeau}]{Mazroui2012}
Mazroui Y, Mathoulin-P\'{e}lissier S, Soubeyran P, Rondeau V (2012).
\newblock \enquote{{General Joint Frailty Model for Recurrent Event Data with a
  Dependent Terminal Event: Application to Follicular Lymphoma Data.}}
\newblock \emph{Statistics in Medicine}, \textbf{31}(11-12), 1162--76.

\bibitem[{Mazroui \emph{et~al.}(2015)Mazroui, Mauguen, Macgrogan,
  Mathoulin-P\'{e}lissier, Brouste, and Rondeau}]{Mazroui2015}
Mazroui Y, Mauguen A, Macgrogan G, Mathoulin-P\'{e}lissier S, Brouste V,
  Rondeau V (2015).
\newblock \enquote{{Time-Varying Coefficients in a Multivariate Frailty Model :
  Application to Breast Cancer Recurrences of Several Types and Death}.}
\newblock \emph{Lifetime Data Analysis}, pp. 1--25.

\bibitem[{Mbogning \emph{et~al.}(2015)Mbogning, Bleakley, and
  Lavielle}]{Mbogning2015}
Mbogning C, Bleakley K, Lavielle M (2015).
\newblock \enquote{Joint Modeling of Longitudinal and Repeated Time-to-Event
  Data Using Nnonlinear Mixed-Eeffects Models and the SAEM Aalgorithm.}
\newblock \emph{Journal of Statistical Computation and Simulation},
  \textbf{85}(8), 1512--28.

\bibitem[{McCrink \emph{et~al.}(2013)McCrink, Marshall, and
  Cairns}]{McCrink2013}
McCrink LM, Marshall AH, Cairns KJ (2013).
\newblock \enquote{{Advances in Joint Modelling: A Review of Recent
  Developments with Application to the Survival of End Stage Renal Disease
  Patients}.}
\newblock \emph{International Statistical Review}, \textbf{81}(2), 249--269.

\bibitem[{Philipson \emph{et~al.}(2012)Philipson, Sousa, Diggle, Williamson,
  Kolamunnage-Dona, and Henderson}]{joineR}
Philipson P, Sousa I, Diggle P, Williamson P, Kolamunnage-Dona R, Henderson R
  (2012).
\newblock \emph{\pkg{joineR}: Joint Modelling of Repeated Measurements and
  Time-to-Event Data}.
\newblock \proglang{R}~package version~1.0-3,
  \urlprefix\url{http://CRAN.R-project.org/package=joineR}.

\bibitem[{Proust-Lima \emph{et~al.}(2016)Proust-Lima, Philipps, Diakite, and
  Liquet}]{lcmm}
Proust-Lima C, Philipps V, Diakite A, Liquet B (2016).
\newblock \emph{\pkg{lcmm}: Extended Mixed Models Using Latent Classes and
  Latent Processes}.
\newblock \proglang{R}~package version~1.7.5,
  \urlprefix\url{http://CRAN.R-project.org/package=lcmm}.

\bibitem[{Proust-Lima \emph{et~al.}(2014)Proust-Lima, S\'{e}ne, Taylor, and
  Jacqmin-Gadda}]{Proust-Lima2014}
Proust-Lima C, S\'{e}ne M, Taylor JMG, Jacqmin-Gadda H (2014).
\newblock \enquote{{Joint Latent Class Models for Longitudinal and
  Time-to-Event Data: A Review.}}
\newblock \emph{Statistical Methods in Medical Research}, \textbf{23}(1),
  74--90.

\bibitem[{Proust-Lima and Taylor(2009)}]{Proust-Lima2009}
Proust-Lima C, Taylor JMG (2009).
\newblock \enquote{{Development and Validation of a Dynamic Prognostic Tool for
  Prostate Cancer Recurrence Using Repeated Measures of Posttreatment PSA: A
  Joint Modeling Approach.}}
\newblock \emph{Biostatistics}, \textbf{10}(3), 535--49.

\bibitem[{Ramsay(1988)}]{Ramsay1988}
Ramsay JO (1988).
\newblock \enquote{{Monotone Regression Splines in Action}.}
\newblock \emph{Statistical Science}, \textbf{3}(4), 425--461.

\bibitem[{{\proglang{R} Development Core Team}(2016)}]{R2016}
{\proglang{R} Development Core Team} (2016).
\newblock \emph{\proglang{R}: A Language and Environment for Statistical
  Computing}.
\newblock \proglang{R} Foundation for Statistical Computing, Vienna, Austria.
\newblock \urlprefix\url{http://www.R-project.org}.

\bibitem[{Rizopoulos(2011)}]{Rizopoulos2011a}
Rizopoulos D (2011).
\newblock \enquote{{Dynamic Predictions and Prospective Accuracy in Joint
  Models for Longitudinal and Time-to-Event Data.}}
\newblock \emph{Biometrics}, \textbf{67}(3), 819--29.

\bibitem[{Rizopoulos(2012)}]{Rizopoulos2012}
Rizopoulos D (2012).
\newblock \emph{{Joint Models for Longitudinal and Time-to-Event Data: With
  Applications in \proglang{R}}}.
\newblock CRC Press.

\bibitem[{Rizopoulos(2016{\natexlab{a}})}]{JM}
Rizopoulos D (2016{\natexlab{a}}).
\newblock \emph{\pkg{JM}: Joint Modeling of Longitudinal and Survival Data}.
\newblock \proglang{R}~package version~1.4-4,
  \urlprefix\url{http://CRAN.R-project.org/package=JM}.

\bibitem[{Rizopoulos(2016{\natexlab{b}})}]{JMbayes}
Rizopoulos D (2016{\natexlab{b}}).
\newblock \emph{\pkg{JMbayes}: Joint Modeling of Longitudinal and Time-to-Event
  Data under a Bayesian Approach}.
\newblock \proglang{R}~package version~0.7-9,
  \urlprefix\url{http://CRAN.R-project.org/package=JMbayes}.

\bibitem[{Rondeau and Gonzalez(2005)}]{Rondeau2005}
Rondeau V, Gonzalez JR (2005).
\newblock \enquote{{\pkg{frailtypack}: A Computer Program for the Analysis of
  Correlated Failure Time Data Using Penalized Likelihood Estimation}.}
\newblock \emph{Computer Methods and Programs in Biomedicine}, \textbf{80},
  154--164.

\bibitem[{Rondeau \emph{et~al.}(2016)Rondeau, Gonzalez, Mazroui, Mauguen, Krol,
  Diakite, and Laurent}]{frailtypack}
Rondeau V, Gonzalez JR, Mazroui Y, Mauguen A, Krol A, Diakite A, Laurent A
  (2016).
\newblock \emph{\pkg{frailtypack}: General Frailty Models: Shared, Joint and
  Nested Frailty Models with Prediction}.
\newblock \proglang{R}~package version~2.9.4,
  \urlprefix\url{http://CRAN.R-project.org/package=frailtypack}.

\bibitem[{Rondeau \emph{et~al.}(2007)Rondeau, Mathoulin-P\'{e}lissier,
  Jacqmin-Gadda, Brouste, and Soubeyran}]{Rondeau2007joint}
Rondeau V, Mathoulin-P\'{e}lissier S, Jacqmin-Gadda H, Brouste V, Soubeyran P
  (2007).
\newblock \enquote{{Joint Frailty Models for Recurring Events and Death Using
  Maximum Penalized Likelihood Estimation: Application on Cancer Events}.}
\newblock \emph{Biostatistics}, \textbf{8}(4), 708--721.

\bibitem[{Rondeau \emph{et~al.}(2012)Rondeau, Mazroui, and
  Gonzalez}]{Rondeau2012frailtypack}
Rondeau V, Mazroui Y, Gonzalez JR (2012).
\newblock \enquote{{\pkg{frailtypack}: An \proglang{R} Package for the Analysis
  of Correlated Survival Data with Frailty Models Using Penalized Likelihood
  Estimation or Parametrical Estimation}.}
\newblock \emph{Journal of Statistical Software}, \textbf{47}(4).

\bibitem[{Rondeau \emph{et~al.}(2011)Rondeau, Pignon, and
  Michiels}]{Rondeau2011joint}
Rondeau V, Pignon JP, Michiels S (2011).
\newblock \enquote{{A Joint Model for the Dependence between Clustered Times to
  Tumour Progression and Deaths: A Meta-Analysis of Chemotherapy in Head and
  Neck Cancer}.}
\newblock \emph{Statistical Methods in Medical Research}.

\bibitem[{Sinha \emph{et~al.}(2008)Sinha, Maiti, Ibrahim, and
  Ouyang}]{Sinha2008}
Sinha D, Maiti T, Ibrahim JG, Ouyang B (2008).
\newblock \enquote{{Current Methods for Recurrent Events Data with Dependent
  Termination: A Bayesian Perspective}.}
\newblock \emph{Journal of American Statistical Association},
  \textbf{103}(482), 866--878.

\bibitem[{Therneau and Lumley(2016)}]{survival}
Therneau TM, Lumley T (2016).
\newblock \emph{\pkg{survival}: Survival Analysis}.
\newblock \proglang{R}~package version~2.39-5,
  \urlprefix\url{http://CRAN.R-project.org/package=survival}.

\bibitem[{Wulfsohn and Tsiatis(1997)}]{Wulfsohn1997}
Wulfsohn MS, Tsiatis AA (1997).
\newblock \enquote{{A Joint Model for Survival and Longitudinal Data Measured
  with Error.}}
\newblock \emph{Biometrics}, \textbf{53}(1), 330--9.

\bibitem[{Yu and Liu(2011)}]{Yu2011}
Yu Z, Liu L (2011).
\newblock \enquote{{A Joint Model of Recurrent Events and a Terminal Event with
  a Nonparametric Covariate Function}.}
\newblock \emph{Statistics in Medicine}, \textbf{30}, 2683--2695.

\bibitem[{Yu \emph{et~al.}(2014)Yu, Liu, Bravata, and Williams}]{Yu2014}
Yu Z, Liu L, Bravata DM, Williams LS (2014).
\newblock \enquote{{Joint Model of Recurrent Events and a Terminal Event With
  Time-Varying Coefficients}.}
\newblock \emph{Biometrical Journal}, \textbf{56}(2), 183--197.

\bibitem[{Zhang \emph{et~al.}(2014)Zhang, Chen, and Ibrahim}]{JMFit}
Zhang D, Chen MH, Ibrahim J (2014).
\newblock \emph{\pkg{JMFit}: A \proglang{SAS} Macro for Joint Models of
  Longitudinal and Survival Data}.
\newblock \urlprefix\url{http://impact.unc.edu/impact7/jmfit}.

\bibitem[{Zhao \emph{et~al.}(2012)Zhao, Liu, Liu, and Xu}]{Zhao2012}
Zhao X, Liu L, Liu Y, Xu W (2012).
\newblock \enquote{{Analysis of Multivariate Recurrent Event Data with
  Time-Dependent Covariates and Informative Censoring}.}
\newblock \emph{Biometrical Journal}, \textbf{54}(5), 585--599.

\end{thebibliography}
\appendix
\section{Summary of the package frailtypack}
\begin{figure}[h]
\begin{center}
	\includegraphics[ trim = 0.5cm 4cm 0.5cm 0cm, clip, width=16cm]{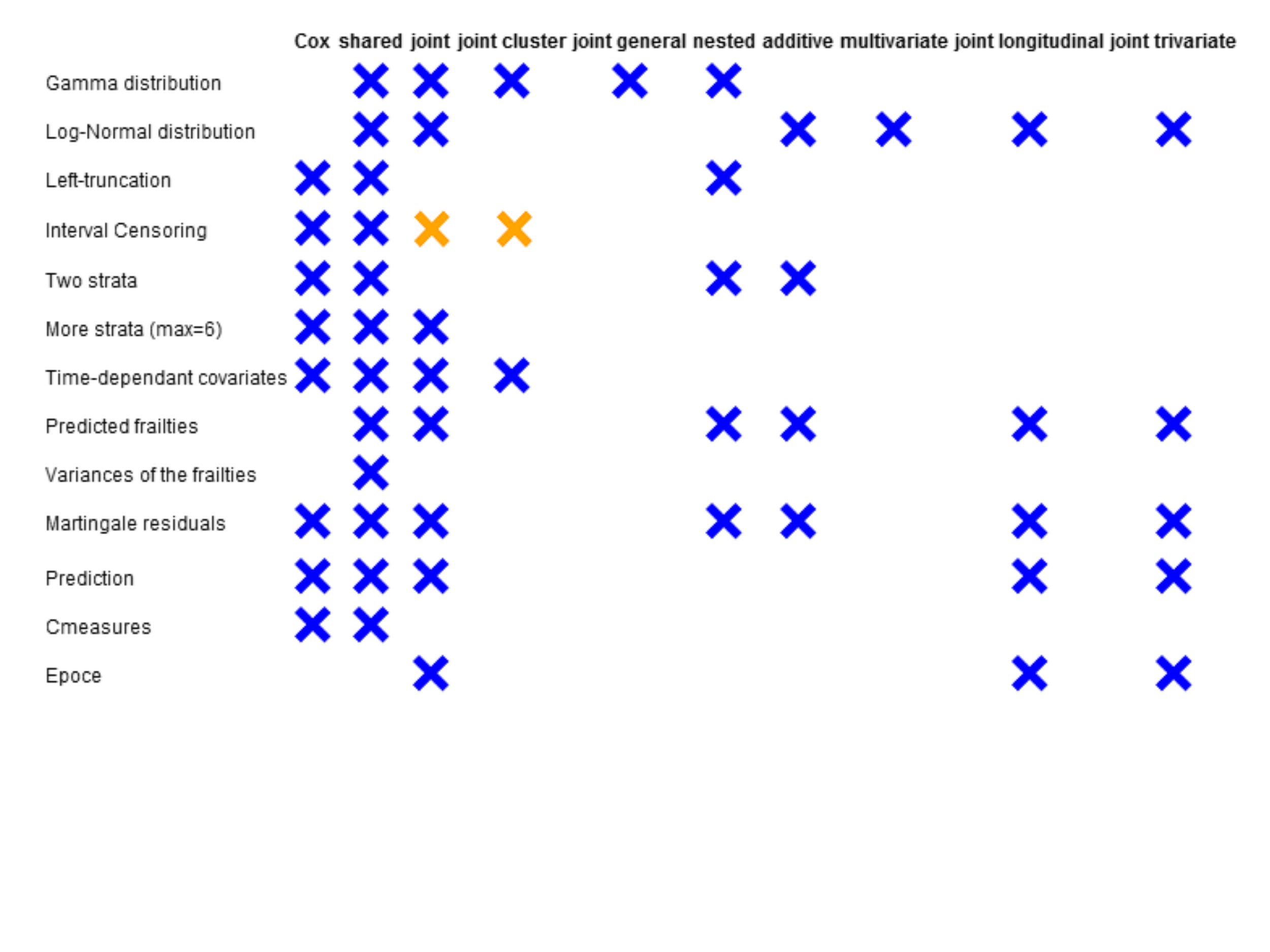} 
	\caption{Package characteristics (version 2.8.3). Blue cross is for option available for a given type of model in the package on CRAN, orange cross is for option included in the package but not on CRAN yet. Empty cells refer to option not available for given types of model. }
	\label{figure:vignette}
	\end{center}
\end{figure}

\section{Additional graphics}
\begin{figure}[h]
\begin{center}
	\includegraphics[ trim = 0.0cm 0.7cm 0.2cm 0.7cm, clip, width=10cm]{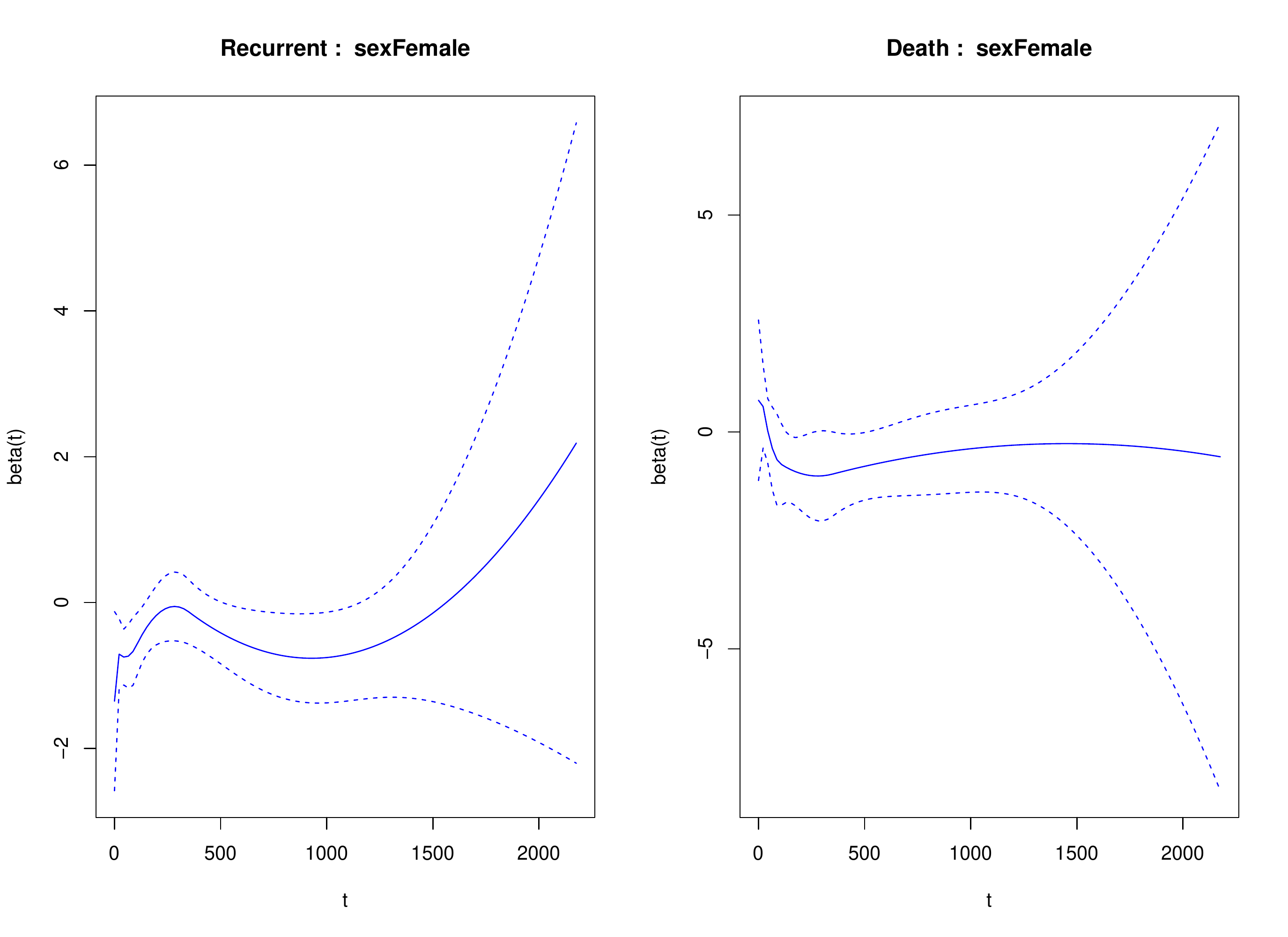} 
	\caption{Log-hazard ratios for sex for \code{modJoint.gap.timedep} model. }
	\label{figure:timedep}
	\end{center}
\end{figure}
\begin{figure}[h!]
\begin{center}
	\includegraphics[ trim = 0.0cm 0.2cm 0cm 0.2cm, clip, width=12cm]{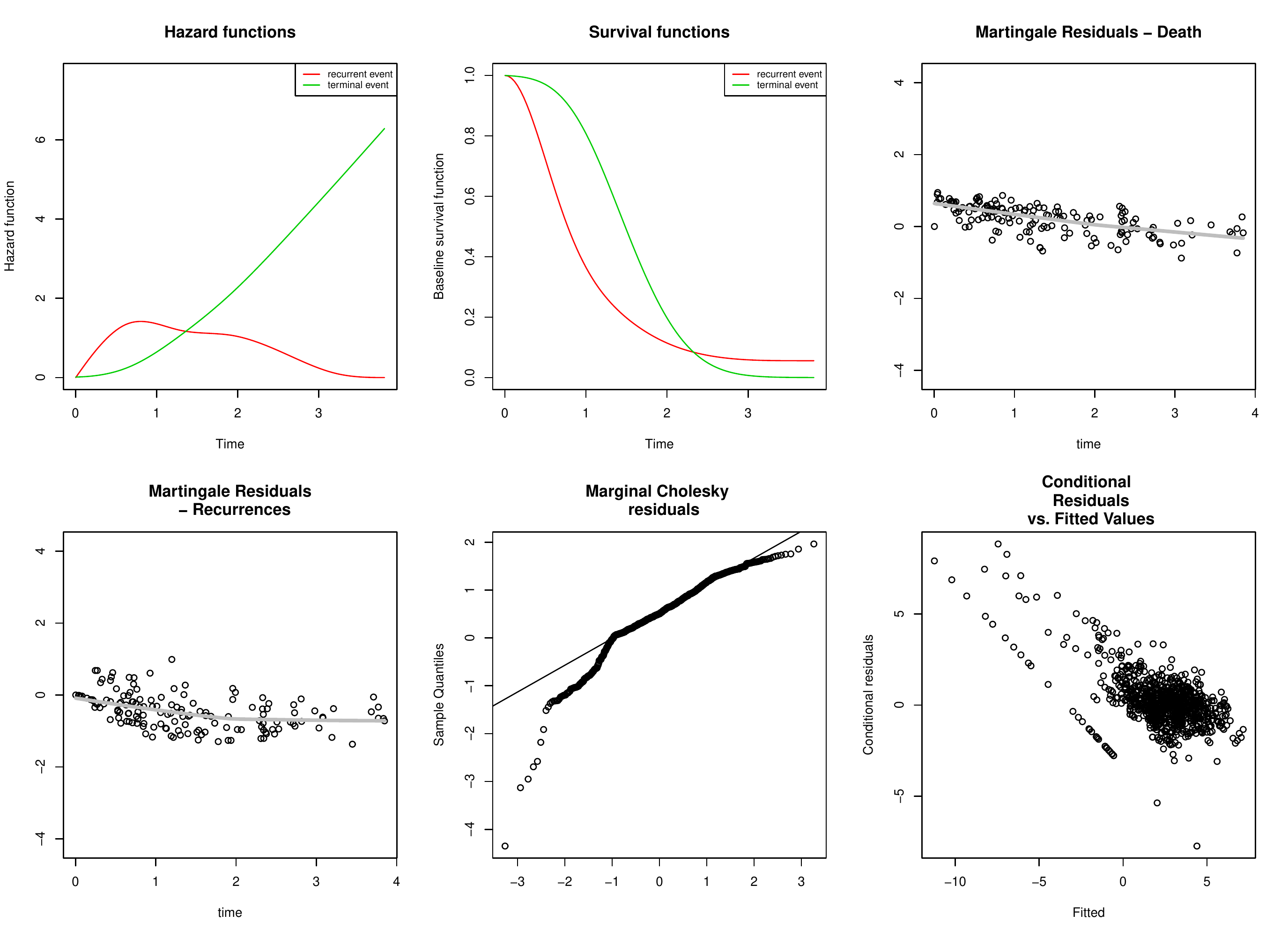} 
	\caption{Goodness-of-fit of trivariate model \code{modTrivariate}. From the top-left plot: baseline hazard and survival functions,  martingale residuals (the grey line corresponds to a smooth curve obtained with \code{lowess}), Q-Q normal plot of the marginal Cholesky residuals of the biomarker and the conditional residuals against the fitted values of the biomarker.}
	\label{figure:residuals_triv}
	\end{center}
\end{figure}

\vspace{10cm}
$\ $

\end{document}